\documentclass[twocolumn, tighten]{aastex63}
\pdfoutput=1
\usepackage{mathtools}
\usepackage{gensymb}
\usepackage{appendix}
\usepackage{bm}
\usepackage[section]{algorithm}
\usepackage[noend]{algpseudocode}

\received{February 28, 2020}
\revised{June 9, 2020}
\accepted{July 22, 2020}
\submitjournal{The Astrophysical Journal}

\begin{document}

\title{Synthetic Spectra of Rotating Stars}

\correspondingauthor{Mikhail Lipatov}
\email{mikhail@physics.ucsb.edu}

\author[0000-0001-9939-1758]{Mikhail Lipatov}
\affiliation{Department of Physics \\
University of California, Santa Barbara \\
Santa Barbara, CA 93106, USA}

\author{Timothy D. Brandt}
\affiliation{Department of Physics \\
University of California, Santa Barbara \\
Santa Barbara, CA 93106, USA}

\begin{abstract}
Many early-type stars have oblate surfaces, spatial temperature variations, and spectral line broadening that indicate large rotational velocities. Rotation ought to have a significant effect on the full spectra of such stars. To infer structural and life history parameters from their spectra, one must integrate specific intensity over the two-dimensional surfaces of corresponding stellar models. Toward this end, we offer PARS (Paint the Atmospheres of Rotating Stars) -- an integration scheme based on models that incorporate solid body rotation, Roche mass distribution, and collinearity of gravity and energy flux \citep{lipatov2020}. The scheme features a closed-form expression for the azimuthal integral, a high-order numerical approximation of the longitudinal integral, and a precise calculation of surface effective temperature at rotation rates up to 99.9\% of Keplerian limit. Extensions of the scheme include synthetic color-magnitude diagrams and planetary transit curves.
\end{abstract}


\section{Introduction} \label{sec:intro}

Advances in optical interferometry over the past two decades enable the resolution of nearby stellar surfaces \citep{monnier_2003_rpp,zhao_2011_iaus,vanbelle_2012_aapr}. Corresponding observations reveal that at least 4 of the 15 brightest early-type stars have non-spherical shapes and star-scale variation in surface temperature: Vega \citep[][henceforth YP10]{yoon_2010}, Achernar \citep{desouza_2014_A&A}, Altair \citep[][henceforth BD20]{bouchaud_2020_A&A}, and Regulus \citep{che_2011_apj}.  Rapid rotation can explain these effects. The polar regions of a spinning star are closer to its core and thus hotter than its equatorial regions \citep{owocki_1994_ApJ,vonzeipel_1924_mras665,cranmer_1995_ApJ}. The Keplerian velocity of such a star provides an upper limit to its surface rotation rate \citep[e.g.,][]{ekstrom_2008_A&A}. Vega, Achernar, Altair, and Regulus all have inferred rotation rates between 0.62 and 0.84 of the Keplerian limit, and polar temperatures 23\% to 35\% hotter than their equatorial temperatures. Here and in the rest of this article, Vega's parameters are from Table 1 with horizontal macroturbulence in YP10, Achernar's are from \citet{desouza_2014_A&A}, Altair's are from Table 5 in BD20, and Regulus's are from Table 4 in \citet{che_2011_apj} with the modified von Zeipel model.

When a star's angular size is too small for interferometric resolution, one can invoke the Doppler effect to infer projected rotation from the broadening of individual spectral lines \citep{elvey_1930_apj,herbig_1955_apj}. Interpretation of observations in light of this methodology supports the view that many unresolved early-type stars rotate at significant fractions of the Keplerian limit \citep{glebocki_2005_ycat,diaz_2011_A&A,Zorec+Royer_2012}. This, in addition to the effect of rotation on the surface temperatures of resolved stars, implies that rapid rotation frequently affects both spectral energy distributions and absorption line profiles of unresolved stars. 

An observer viewing the pole of a spinning star will see a larger, hotter surface in projection than one viewing its equator. Through this effect, stellar rotation can explain color-magnitude diagrams of star clusters with single, coeval populations \citep{brandt_2015_apj24,dejuanovelar_2019_mnras,gossage_2019_apj}. Models of clusters start with the evolution of individual rotating stars from the zero-age main sequence (ZAMS) to the observed epoch. One then needs to integrate specific intensities over the surfaces of stellar models for comparison with the observed spectra, colors, and magnitudes. 

The inhomogeneous surface of a rotating star also affects transit light curves: a misaligned planetary or stellar companion will produce a deeper or shallower transit as it transits a hotter or cooler part of the star \citep{barnes_2009}. Computing this effect from a full stellar model requires calculating both the integrated spectrum and the specific intensity along a chord of finite thickness running across the projected stellar surface. 

This article presents a fast, flexible, and accurate numerical integration scheme to compute synthetic spectra, color-magnitude diagrams, and transit light curves for rapidly rotating stars. The scheme does not take into account the rotational Doppler effect, so that it is mainly applicable to the inference of rotational parameters from broad spectral features, as opposed to specific absorption lines.

We structure the article as follows. Section \ref{sec:model} introduces our stellar model and the double integral that we compute. Section \ref{sec:surface} details the calculation of the stellar surface over which we integrate. Sections \ref{sec:intensity} and \ref{sec:integr} delineate the integration itself. Section \ref{sec:ext} demonstrates extensions of our scheme to compute color-magnitude diagrams and planetary transit curves; we conclude with Section \ref{sec:concl}. 

\section{Stellar model} \label{sec:model}

We begin with our model for the structure of the star itself. A simple choice is a uniformly rotating Roche model, where the potential at the stellar surface is approximated by placing all of the star's mass at its center. However, two-dimensional models of material flows in rotating stars predict that the cores and equatorial regions of early-type stars have larger angular velocities than their envelopes and polar regions, respectively \citep{rieutord_2016_jcoph,rieutord_2009_coast}. BD20 infer these effects from spectro-interferometry of a resolved star. In general, mass flows and differential rotation could result in profound consequences for the shapes and temperatures of stars \citep{kippenhahn_1977_A&A,zorec_2011_A&A}. 

Nevertheless, a Roche model can account for much of the physics that underlies the spectrum of an early-type star. One such model incorporates collinearity of gravity and energy flux \citep[][henceforth ER11]{EspinosaLara+Rieutord_2011}. ER11 show that there is less than 0.01 radians of deviation from this collinearity in a two-dimensional model of material flow. Additionally, the linear dimensions of a polytrope that mimics the star's structure and those of a Roche model differ by about 1\% \citep{orlov_1961_sva}. Furthermore, stellar evolution calculations that form an input to our integration scheme do not model surface differential rotation due to their fundamentally one-dimensional nature. These calculations consider uniform rotation on isobars and use pressure as the radial coordinate (again assuming collinearity of effective gravity and energy flux and uniform surface rotation) \citep{Meynet+Maeder_2000,Ekstrom+Georgy+Eggenberger+etal_2012,georgy_2014_A&A,Paxton+Bildsten+Dotter+etal_2011,Paxton+Smolec+Schwab+etal_2019}. The calculations result in surface shapes and temperature profiles that agree with those of ER11's structural model. Fully two-dimensional stellar evolution models remain well beyond reach due to the enormous range of both time and length scales in the problem.

Due to the match between ER11's model and the output of stellar evolution calculations, its close agreement with fully two-dimensional structural models, and numerical convenience, we adopt ER11's model and note that it involves solid body rotation.

\begin{figure}
\includegraphics[width=0.9\linewidth]{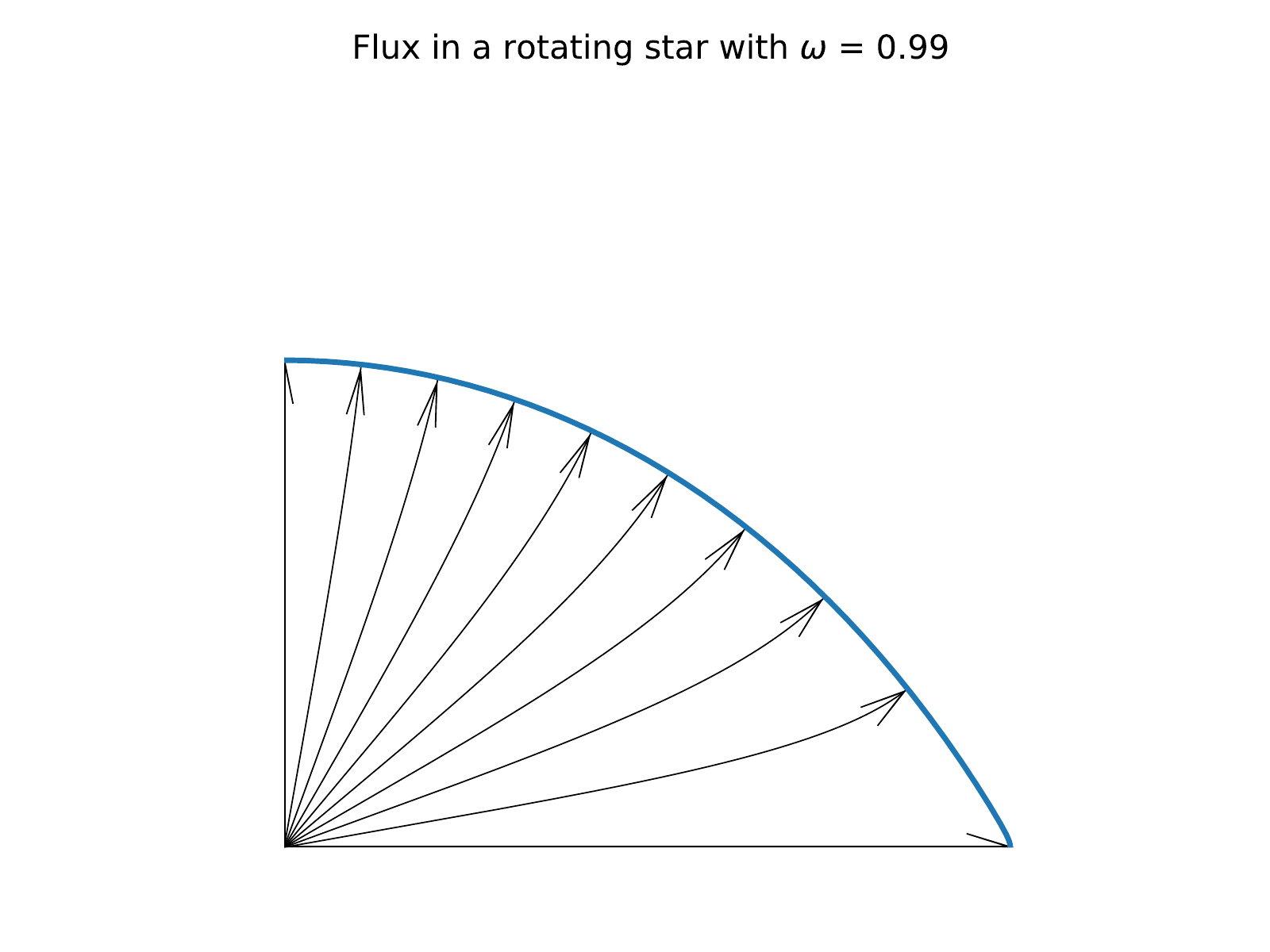}
\caption{ Meridional cut through the projected surface and internal energy flux lines of a star rotating at $\omega = 0.99$, from ER11's model. Only 1/4 of the star is shown, due to symmetry. At the center of the star, the flux lines are equally spaced. At the surface, however, they are farther apart near the equator than near the pole (the pole, as a result, is hotter).}
\label{fig:flux}
\end{figure}

We can visualize energy transport through a rotating star by solving for the internal energy flux lines.  Figure \ref{fig:flux} shows these lines as computed numerically from ER11's equation 21.  They are relatively far from each other in the equatorial regions of the surface, so that these regions are at relatively low temperatures due to the Stefan-Boltzmann law.  At the star's center, where rotation is dynamically unimportant (under the assumption of solid body rotation), the flux lines are equally spaced and energy transport is independent of polar angle.

\begin{figure}
\centering
\includegraphics[width=0.8\linewidth]{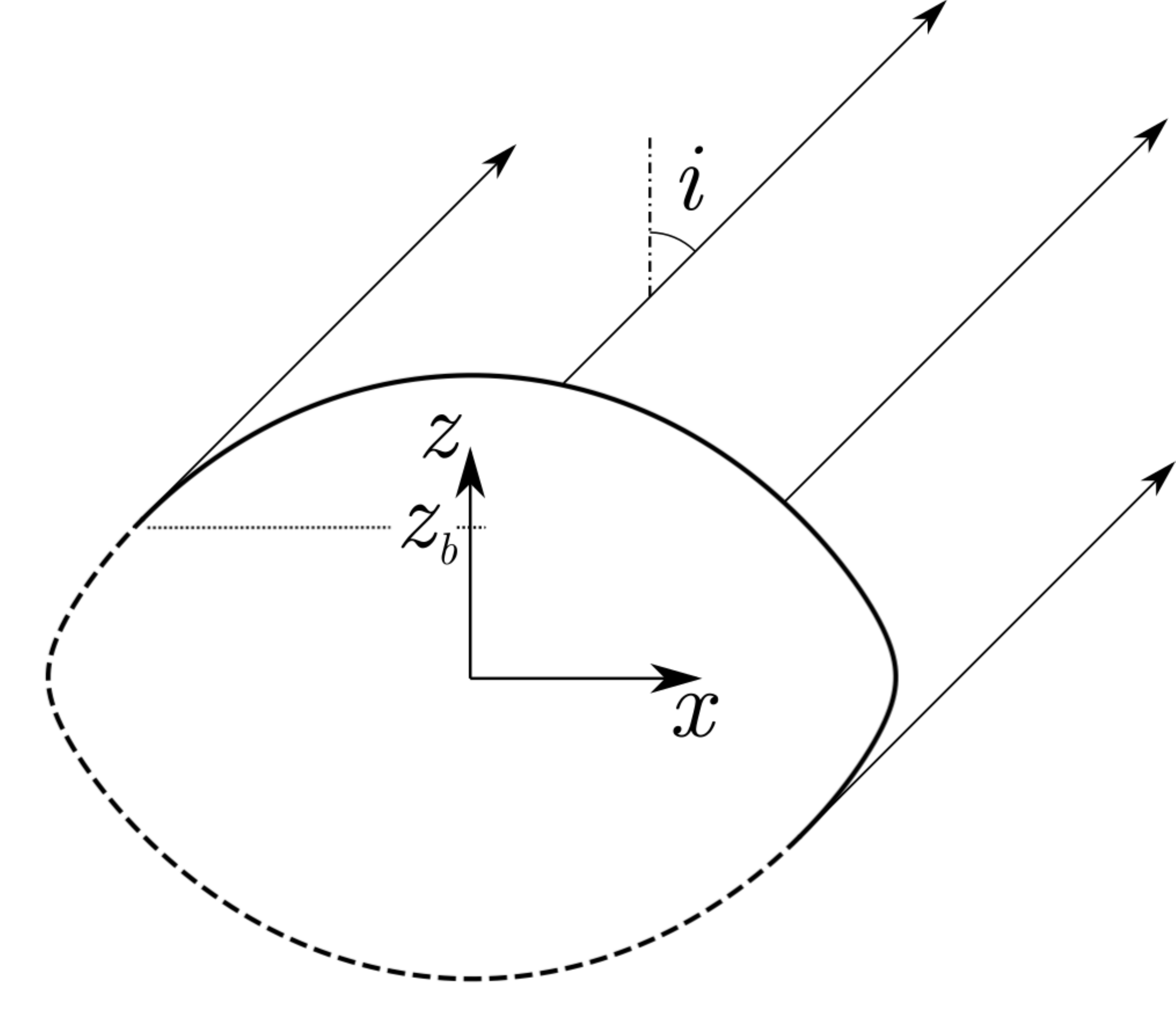}
\caption{ Meridional cut of the projected surface and observer sightlines for a star rotating at 90\% of its Keplerian surface velocity with $i = \pi/4$ (see Section \ref{sec:model}). The star is visible at all $\phi$ above $z=z_{b}$ and, by symmetry, at no $\phi$ below $z=-z_b$.}
\label{fig:sight}
\end{figure}

We define inclination $i \in [0, \pi/2]$ as the angle between the model's rotation axis $\mathbf{\hat{z}}$ and the line of sight $\mathbf{\hat{i}}$. The ER11 model is symmetric about its axis of rotation, though not necessarily about any line of sight. Accordingly, we primarily describe the stellar surface by a set of cylindrical coordinates $z$, $r$ and $\phi$ defined by the stars's rotational symmetry and assign $\phi = 0$ to the azimuthal direction closest to $\mathbf{\hat{i}}$. We also use Cartesian coordinates that include $\mathbf{\hat{z}}$ and an $x$-axis coinciding with $\phi = 0$ at $z = 0$ (see Figure \ref{fig:sight}). We define $R_p$ as the polar radius of the star and $\mu \in [0, 1]$ as cosine of the viewing angle, i.e. the angle between the line of sight and the normal to the surface.

Our goal is to compute the star's specific flux $\mathcal{F}_\nu$ along the line of sight. $\mathcal{F}_\nu$ is given by
\begin{equation}
D_{\star}^2\, \mathcal{F}_\nu = 2 \int^{R_p}_{-z_b} A(z) \, \int^{\phi_b(z)}_{0} I_\nu(z, \phi)\,\mu(z, \phi)\,d\phi\,dz,
\label{eq:int}
\end{equation}
where $D_{\star}$ is the distance to the star and $z_b$ corresponds to the lowest stellar latitude that is visible at all $\phi$. For a differential element of the stellar surface, $A(z)\,\mu(z, \phi)\,d\phi\,dz$ is the element's area projected onto the view plane and $I_\nu(\phi, z)$ is its specific intensity per unit projected area.
For all $z > -z_b$, at least some of the star is visible, while none of it is visible for $z < -z_b$. At a given $z$, the star is visible everywhere between $-\phi_b(z)$ and $\phi_b(z)$ and is not visible anywhere else. The factor of 2 arises due to the symmetries of the model, since the integral between 0 and $\phi_b(z)$ is equal to that between $-\phi_b(z)$ and 0. Another result of the symmetries is the fact that $\phi_b(z) = \pi$ for $z > z_b$. 

\section{Visible surface} \label{sec:surface}

\subsection{Surface shape} \label{subsec:shape}

We define $\tilde{r}\equiv r / R_e$, $\tilde{z}\equiv z / R_p$, and flatness $f \equiv R_e / R_p$, where $R_e$ is the star's equatorial radius. Our definition of $\tilde{r}$ is different from that in ER11, which reserves this symbol for a normalized spherical coordinate. Note that $r'(z) = f\,\tilde{r}'(\tilde{z})$. By substituting the polar value of ER11's normalized spherical coordinate into their Equation (30), we find that 
\begin{equation}
f = 1 + \omega^2 / 2,
\label{eq:f}
\end{equation}
where 
\[
\omega \equiv \Omega \sqrt{\frac{R_e^3}{G M}} = \frac{\Omega}{\Omega_k},
\]
$\Omega$ is the star's angular velocity, $\Omega_k$ is the Keplerian velocity, $M$ is the mass of the star, and $G$ is the gravitational constant. 

We define the following helper variables and constants:
\begin{equation}
w \equiv 1 + 2 / \omega^2,\,\, u \equiv \tilde{z} / f = z / R_{e} \,\, \mathrm{and}\,\, s \equiv \tilde{r}^2.
\label{eq:wus}
\end{equation}
In the remainder of this article, we will indicate point locations by either dimensional coordinates such as $z$ or by normalized dimensionless coordinates such as $\tilde{z}$ or $u$. Converting ER11's Equation (30) to normalized cylindrical coordinates and keeping in mind Equation \eqref{eq:wus}, we obtain
\begin{equation}
\frac{1}{\omega^2 \sqrt{\tilde{r}^2 + u^2}} + \frac{\tilde{r}^2}{2}= \frac{1}{\omega^2} + \frac{1}{2},
\label{eq:surf}
\end{equation}
which leads to a cubic in $s$:
\begin{multline}
s^3+s^2 \left(u^2-2 w\right)+s\, w \left(w-2 u^2\right)+\\
\left(u^2-1\right) w^2+2 w-1 = 0.
\label{eq:cubic}
\end{multline}
We solve this as a function of $u$ to obtain
\begin{equation}
s(u) = \frac{1}{3} \left[\, -u^2 + 2 w + 2 \left(u^2+w\right) v(u) \,\right].
\label{eq:s}
\end{equation}
Here,
\begin{equation}
v(u) \equiv \cos \left[\frac{1}{3} \left(\cos ^{-1}\left[\,t(u)\,\right]+2 \pi \right)\right]
\label{eq:v}
\end{equation}
with
\begin{equation}
t(u) \equiv \frac{27\,(1 - w)^2}{2 \left(u^2+w\right)^3} - 1.
\label{eq:t}
\end{equation}
We differentiate Equation \eqref{eq:s} to obtain
\begin{equation}
s'(u) = \frac{2 u}{3}\, \frac{1 - 2\,v(u)}{1 + 2\,v(u)}.
\label{eq:Ds}
\end{equation}
Equations \eqref{eq:f}--\eqref{eq:Ds}, along with the definitions of $f$, $\tilde{r}$, and $\tilde{z}$, can be used to obtain $\tilde{r}(\tilde{z})$ and $\tilde{r}'(\tilde{z})$. Now consider
\begin{equation}
\mathbf{dl} \equiv \mathbf{\hat{r}}\,dr + \mathbf{\hat{z}}\,dz,
\label{eq:dlvec}
\end{equation}
a differential element of $r(z)$ between $z$ and $z + dz$. Its length $dl$ can be found from the Pythagorean theorem and the definition of $r'(z)$: 
\begin{equation}
dl = \sqrt{r'(z)^2 + 1}\, dz.
\label{eq:dl}
\end{equation}
As we rotate $\mathbf{dl}$ around the star’s symmetry axis by $d\phi$, the area of the resulting differential surface element is $r(z)\,dl\,d\phi$. We multiply this area by $\mu(z, \phi)$ to project it onto the view plane, substitute for $dl$ according to Equation \eqref{eq:dl}, and change variables from $z$ and $r(z)$ to $\tilde{z}$ and $\tilde{r}(\tilde{z})$. This results in
\begin{equation}
A(z)\,\mu(z, \phi)\,dz\,d\phi = R_{e}^2\,\tilde{A}(\tilde{z})\,\mu(\tilde{z}, \phi)\, d\tilde{z}\,d\phi
\label{eq:area1}
\end{equation}
with
\begin{equation}
\tilde{A}(\tilde{z}) = \frac{\tilde{r}(\tilde{z})\, n(\tilde{z})}{f},
\label{eq:area3}
\end{equation}
where
\begin{equation}
n(\tilde{z}) = \sqrt{\left[f\,\tilde{r}'(\tilde{z})\right]^2 + 1}.
\label{eq:n}
\end{equation}
A change of variables from $\tilde{r}$ and $\tilde{z}$ to $s$ and $u$ implies
\begin{equation}
f\, \tilde{r}'(\tilde{z}) = \left.\frac{s'(u)}{2\sqrt{s(u)}}\right|_{u = \tilde{z}/f}
\label{eq:rprime}
\end{equation}
and
\begin{equation}
s'(u) = \left. 2 f\, \tilde{r}'(\tilde{z})\, \tilde{r}(\tilde{z})\,\right|_{\tilde{z} = fu}.
\label{eq:rprime2}
\end{equation}
Equation \eqref{eq:rprime2} helps perform this change of variables in equations \eqref{eq:area3} and \eqref{eq:n}. The result is
\begin{equation}
\tilde{A}(\tilde{z}) = \left. \frac{1}{f}\sqrt{\frac{1}{4} s'(u)^2 + s(u)}\right|_{u = \tilde{z}/f}.
\label{eq:area4}
\end{equation}
At the poles of the star, $\tilde{r}(\tilde{z}) \to 0$ and $\tilde{r}'(\tilde{z}) \to \mp \infty$ as $\tilde{z} \to \pm 1$. Thus, the right side of equation \eqref{eq:area3} multiplies zero by infinity in this limit, and we cannot use it to find $\tilde{A}(\pm 1)$, which is finite. However, $s'(u)$ remains finite as $u \to \pm 1/f$. Accordingly, we use equation \eqref{eq:area4} for all computation.

We now re-write Equation \eqref{eq:int} as
\begin{equation}
D_{\star}^2\, \mathcal{F}_\nu = 2 R_{e}^2 \int^{1}_{-\tilde{z}_b} \tilde{A}(\tilde{z}) \, \int^{\phi_b(\tilde{z})}_{0} \mu(\tilde{z}, \phi)\, I_\nu(\tilde{z}, \phi)\,d\phi\,d\tilde{z},
\label{eq:int0}
\end{equation}
where $\tilde{z}_b \equiv z_b / R_p$.

\subsection{Viewing angle}
Due to the star's cylindrical symmetry, a normal to its surface does not have a component in the $\mathbf{\hat{\phi}}$ direction. Such a normal also has to be perpendicular to $\mathbf{dl}$ in Equation \eqref{eq:dlvec}. The two directions that satisfy both conditions are along vectors $\mathbf{dn} = \pm (\mathbf{\hat{r}}\,dz - \mathbf{\hat{z}}\,dr)$.
%One can check that $\mathbf{dn \cdot dl} = 0$.
Here, the positive sign gives the vector with a non-negative $\mathbf{\hat{r}}$ component --- the vector that points away from the star's interior. We divide it by $dz$ to get $\mathbf{\hat{r}} - r'(z) \,\mathbf{\hat{z}}$, make the same change of variables as in Equation \eqref{eq:area1}, and normalize the result, which yields 
\begin{equation}
\mathbf{\hat{n}} = \frac{\mathbf{\hat{r}}-f \, \tilde{r}'(\tilde{z})\,\mathbf{\hat{z}}}{n(\tilde{z})}.
\end{equation}
To convert this expression to the Cartesian coordinates, we replace $\mathbf{\hat{r}}$ with $\cos{\phi}\,\,\mathbf{\hat{x}} + \sin{\phi}\,\,\mathbf{\hat{y}}$:
\begin{equation}
\mathbf{\hat{n}} = \frac{\cos{\phi}\,\,\mathbf{\hat{x}} + \sin{\phi}\,\,\mathbf{\hat{y}} -f \, \tilde{r}'(\tilde{z})\,\mathbf{\hat{z}}}{n(\tilde{z})}.
\end{equation}
The normalized line-of-sight vector is
\begin{equation}
\mathbf{\hat{i}} = \sin{i}\,\,\mathbf{\hat{x}} + \cos{i}\,\,\mathbf{\hat{z}},
\end{equation}
so that the cosine of the angle between the two vectors is
\begin{equation}
\mathbf{\hat{n}} \bm{\cdot} \mathbf{\hat{i}} \equiv \mu(\tilde{z}, \phi) = \frac{\sin{i}\,\,\cos{\phi} - \cos{i}\,\left[f \, \tilde{r}'(\tilde{z})\right]}{n(\tilde{z})}.  
\label{eq:mu}
\end{equation}

\subsection{Visibility Boundaries} \label{subsec:vis}

Setting Equation \eqref{eq:mu} to zero at $\phi = \pi$, we obtain a condition for $\tilde{z} = \tilde{z}_b$ (see Section \ref{sec:model} and Figure \ref{fig:sight}): 
\begin{equation}
f\, \tilde{r}'(\tilde{z}_b) = -\tan{i}.
\label{eq:zb}
\end{equation}
Combining Equations \eqref{eq:zb} and \eqref{eq:rprime} and squaring the result, we obtain
\begin{equation}
\imath \equiv (\tan{i})^2 = \frac{s'(u_b)^2}{4\,s(u_b)},
\label{eq:tani}
\end{equation}
where $u_b \equiv \tilde{z}_b / f$. We solve Equation \eqref{eq:s} for $v(u)$ in terms of $s(u)$ and substitute the result into Equation \eqref{eq:Ds}, thus obtaining $s'(u)$ in terms of $s(u)$. This, in turn, is substituted into equation \eqref{eq:tani}, which results in

\begin{multline}
9 s^3 \imath - s^2 \left[12 u^2 \imath - u^2 - 6 \imath w\right]+\\
s \left[\left(2 u^2 - w \right)^2 \imath + 2 u^2 w^2 \right] -
u^2 w^2 = 0,
\label{eq:sb_ub}
\end{multline}
where, for compactness, $u_b$ is written as $u$ and $s(u_b)$ as $s$. We can solve Equation \eqref{eq:cubic} for $u^2$ in terms of $s(u)$ and substitute the result into Equation \eqref{eq:sb_ub} to obtain a 7th-order polynomial equation in $s(u_b)$, which can be solved numerically. Equation \eqref{eq:cubic} can then be used again to obtain $u_b^2$, $u_b$ and $\tilde{z}_b$.

As discussed in Section \ref{sec:model}, the surface is visible at a subset of $\phi$ for $\tilde{z} \in (-\tilde{z}_b, \tilde{z}_b)$. At every $\tilde{z}$ in this region, there is some $\phi_b$ for which $\mu = 0$. Accordingly, to find these boundary values, we substitute $\phi = \phi_b$ into Equation \eqref{eq:mu}, obtaining
\begin{equation}
\phi_b(\tilde{z}) = \arccos{\left[\,f\, \tilde{r}'(\tilde{z})\,\cot{i}\,\right]}.
\label{eq:phib}
\end{equation}

\section{Azimuthal integral} \label{sec:intensity}

\subsection{Intensity functions} \label{subsec:fits}

\citet{castelli_2004}, henceforth CK04, provide specific intensities $I_\nu$ on a discrete grid of microturbulent velocity $\xi$, metallicity ${\rm [M/H]}$, effective surface temperature $T$, effective surface gravity $g$, radiation wavelength $\lambda$, and cosine of the viewing angle $\mu$. The unit of $I_\nu$ is $\mathrm{erg\,s^{-1}\,Hz^{-1}\,ster^{-1}\,cm^{-2}}$; fixed sets of 1221 $\lambda$ values and 17 $\mu$ values constitute the grid's extent in the corresponding dimensions at all points. Hereafter, effective surface gravity and effective surface temperature are sometimes simply gravity and temperature. 

We define a set of closed intervals $\{m_j\}$ that form a partition of $\mu$'s range, with each interval boundary point among the constant $\mu$ grid values in CK04's data. With $m_j \equiv [\mu_j, \mu_{j+1}]$, $\mu_{j_1} < \mu_{j_2}$ when $j_1 < j_2$.

In the remainder of this work, we set the partition to $[0, 0.1]$, $[0.1, 0.4]$, and $[0.4, 1]$; we also set $[{\rm M / H}]$ to $-0.1$ and $\xi$ to $2\,{\rm km\,s^{-1}}$. Let us say that CK04 provide intensity $I_1$ at $\mu = 1$, as well as intensity values at other discrete $\mu$, all for a specific parameter space location $(T, g, \lambda)$. We model $I_\nu(\mu)$ at this location as a piecewise polynomial
\begin{equation}
I_\nu(\mu) = \sum^{4}_{i=0} a_{ik} \mu^i\quad{\rm with}\,\,\, k = \max_{\mu \in m_{j}} j. 
\label{eq:poly}
\end{equation}
For each $j$, we obtain the coefficients $a_{ij}\,\,\forall i$ by a least-squares fit to CK04's points on $m_j$, using every boundary point on each of the two intervals it belongs to. Here and elsewhere in this article, we use version 3 of the Python programming language and the NumPy library \citep{numpy} for all calculations.

For every $(T, g, \lambda)$, we conduct the above fitting procedure and calculate the associated error in $I_\nu(\mu)$ at every $\mu$ grid point:
\begin{equation}
\left|\frac{\delta I_\nu}{I_1}\right| \equiv \left|\frac{I_\nu(\mu) - I_\mu}{I_1}\right|.
\label{eq:erI}
\end{equation}
Here, $I_\mu$ is CK04's value of $I_\nu$ at the grid point, $I_\nu(\mu)$ is given by equation \eqref{eq:poly}, and the error is normalized by $I_1$. 

\begin{figure}
\includegraphics[width=\linewidth]{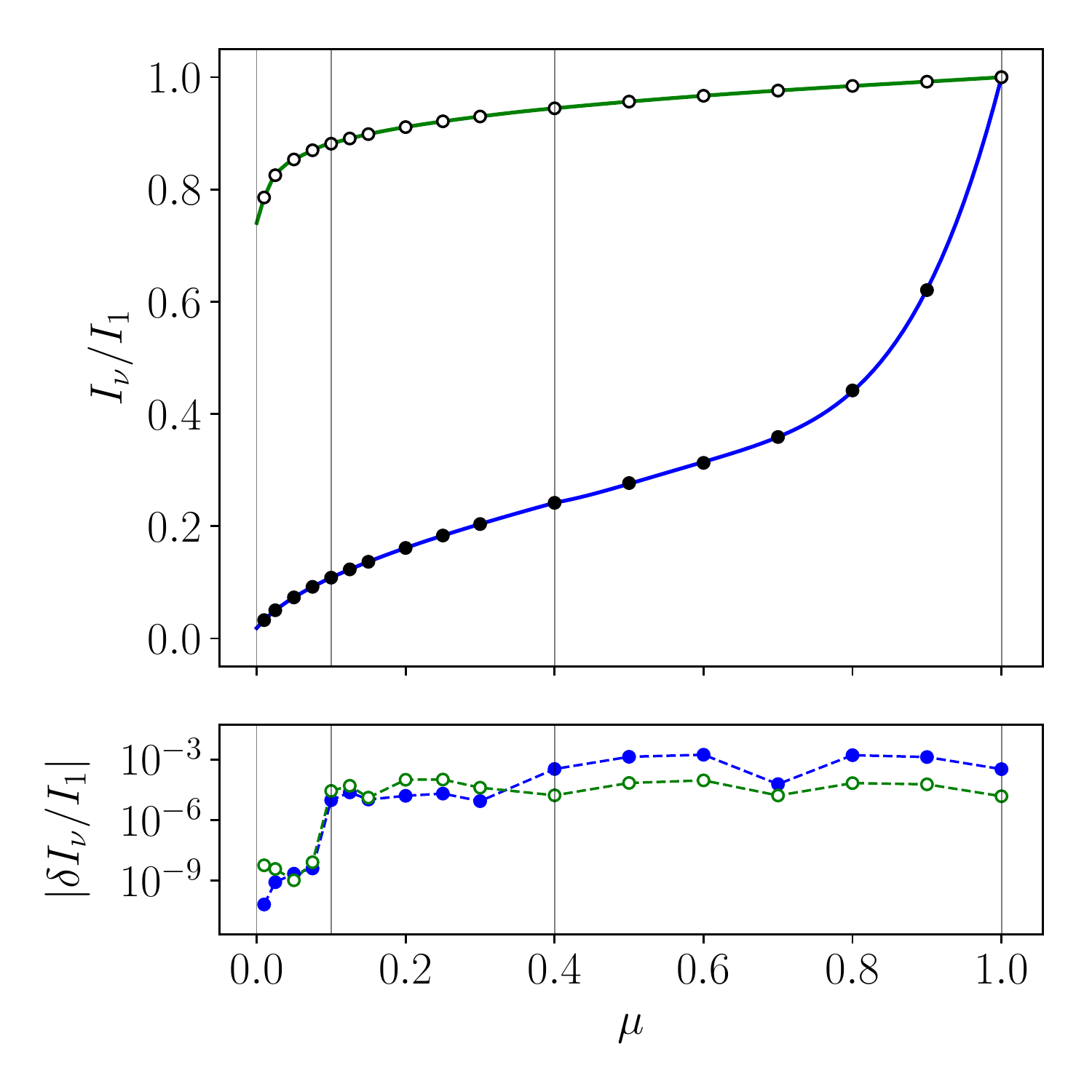}
\caption{Top panel: circles indicate normalized intensity values in CK04 on a grid of $\mu$, lines --- the piecewise polynomial fits $I_\nu(\mu)$ to these points. Bottom panel: relative error in $I_\nu$ at the grid points (see Section \ref{subsec:fits}). Both panels: blue lines and solid markers correspond to the location of maximum error in parameter space: $T = 9000\,{\rm K}$, $\log_{10}{g} = 3.0$, and $\lambda = 111.5\,{\rm nm}$; green lines and open markers --- a location of an error closest to median: $T = 7750\,{\rm K}$, $\log_{10}{g} = 1.5 $, and $\lambda = 5070\,{\rm nm}$; grey vertical lines mark the boundaries of the partition.}
\label{fig:erI}
\end{figure}

The global error maximum and the median of the error maxima across the $(T, g, \lambda)$ space are 0.17\% and 0.010\%, respectively. Figure \ref{fig:erI} presents both the errors and the intensity fits for the location of the global maximum and for one of the locations with maximum error closest to the median. Left to right, the three partition intervals respectively contain 5, 7 and 7 grid points. Thus, the 4th degree polynomial fit is slightly over-constrained on each of the latter two intervals, so that their errors in Figure \ref{fig:erI} give us a sense of the true error associated with the procedure. On the other hand, the lowest interval's number of grid points equals the number of the polynomial's parameters, so that its error is due to round-off.

The lowest $I_\nu(\mu)\,/\,I_1$ and $I'_\nu(\mu)\,/\,I_1$ across all the fits are -0.95\% and -2.65, respectively. Negative values for both quantities are not physical, though they are also rare. We do not expect them to affect the accuracy of our results any more significantly than the errors we estimate via the fitting procedure.

The least-squares fits over the entire parameter space take about 50 seconds on a 2.3 GHz MacBook Pro with 8 GB of RAM. In the remainder of this article, the distinction between $i$ as either an integer-valued index or the inclination should be clear from context. 

\subsection{Piecewise integration} \label{subsec:piecewise}
In the foregoing, $i \in \{0,\ldots,4\}$ and $j \in \{1,2,3\}$. We define functions
\begin{equation}
p_{ij}(\tilde{z}, \phi) \equiv 
\begin{cases} 
    \mu(\tilde{z}, \phi)^i & \mu(\tilde{z}, \phi) \in m_j \\
    0 & \mathrm{otherwise}
\end{cases}.
\label{eq:p}
\end{equation}
Here, $\mu\left(\tilde{z}, \phi\right)$ is given by Equation \eqref{eq:mu}. We re-write Equation \eqref{eq:poly} as
\begin{equation}
I_\nu(\tilde{z}, \phi) = \sum_{i,j}\, a_{ij}(\tilde{z})\,\, p_{ij}(\tilde{z}, \phi),
\label{eq:poly2}
\end{equation}
where the $a_{ij}$ depend on $\tilde{z}$ because $T$ and $g$ depend on $\tilde{z}$ (see Sections \ref{subsec:fits} and \ref{subsec:tgcalc}). We substitute Equation \eqref{eq:poly2} into Equation \eqref{eq:int0} and move both the sum and the fit coefficients outside the integral in $\phi$. This results in
\begin{multline}
D_{\star}^2\,\mathcal{F}_\nu = 2 R_{e}^2 \int^{1}_{-\tilde{z}_b} \tilde{A}(\tilde{z}) \sum_{i,j} a_{ij}(\tilde{z}) \times \\
\int^{\phi_b(\tilde{z})}_{0} \mu(\tilde{z}, \phi)\, p_{ij}(\tilde{z}, \phi)\,d\phi\,d\tilde{z}.
\label{eq:int2}
\end{multline}
We define 
\begin{equation}
P_{ij}(\tilde{z}) \equiv \int^{\phi_b(\tilde{z})}_{0} \mu(\tilde{z}, \phi)\, p_{ij}(\tilde{z}, \phi)\,d\phi.
\label{eq:P}
\end{equation}

According to equation \eqref{eq:p}, every $p_{ij}$ is zero outside $m_j$ and its dependence on $\phi$ is polynomial in $\cos{\phi}$ within $m_j$. Due to this choice of $\{p_{ij}\}$, we can analytically express the indefinite version of each integral in equation \eqref{eq:P} in terms of cosines and sines. Calculation of the definite integrals involves Algorithm \eqref{alg:piecewise}, which keeps track of $m_j$ in the course of integration and accounts for the fact that $\mu(\tilde{z}, \phi)$ in equation \eqref{eq:mu} decreases as $\phi$ increases. Expression of the indefinite integrals in terms of cosines and sines permits quick calculation of $\{P_{ij}\}$. Such calculation may not be possible for a form of $I_\nu$ that differs from equations \eqref{eq:poly}--\eqref{eq:poly2}. For example, a number of authors use forms with $\mu^{1/2}$ \citep[][BD20]{claret_2000_A&A, claret_2018_A&A}, which can be emulated with $i = 1/2$ in some of the $\{p_{ij}\}$. Expressions for the corresponding $\{P_{ij}\}$ involve incomplete elliptic integrals, which are relatively slow to evaluate.

Together, Equations \eqref{eq:int2} and \eqref{eq:P} yield
\begin{equation}
D_{\star}^2\,\mathcal{F}_\nu = 2 R_{e}^2 \int^{1}_{-\tilde{z}_b} \tilde{A}(\tilde{z}) \sum_{i,j} a_{ij}(\tilde{z})\, P_{ij}(\tilde{z})\,d\tilde{z}.
\label{eq:int3}
\end{equation}

\subsection{Temperature and gravity calculation} \label{subsec:tgcalc}

We add spherical coordinates $\theta$ and $\rho$ to our description of the stellar surface. These satisfy
\begin{equation}
\rho = \sqrt{r^2 + z^2}\,\,\, \mathrm{and}\,\,\sin{\theta} = r / \rho.
\end{equation}
We further define $\tilde{\rho}\equiv \rho / R_e$, so that
\begin{equation}
\tilde{\rho} = \sqrt{\tilde{r}^2 + \tilde{z}^2 / f^2}\,\,\, \mathrm{and}\,\,\sin{\theta} = \tilde{r} / \tilde{\rho}.
\label{eq:rho_theta}
\end{equation}
Combined with ER11's Equation (31), this gives us the following expression for gravity:
\begin{equation}
g(\tilde{z}) = \frac{G M}{R_{e}^2} \sqrt{ \frac{1}{\tilde{\rho}^4} + \omega^2 \tilde{r}^2 \left(\omega^2 - \frac{2}{\tilde{\rho}^3}\right)},
\label{eq:g}
\end{equation}
where $\tilde{r}(\tilde{z})$ is found in Section \ref{sec:surface} and $\tilde{\rho}(\tilde{z})$ can be obtained from Equation \eqref{eq:rho_theta}. According to ER11's equations 31 and 26, temperature is then
\begin{equation}
T(\tilde{z}) = \left[\frac{L}{4 \pi \sigma G M} \, F(\tilde{z})\,  g(\tilde{z})\right]^{1/4},
\label{eq:T}
\end{equation}
where $L$ is the luminosity of the star, $\sigma$ is Stefan's constant and 
\begin{equation}
F = \left(\frac{\tan{\vartheta}}{\tan{\theta}}\right)^2
\label{eq:F}
\end{equation}
with
\begin{equation}
\cos{\vartheta} + \ln{\tan{\frac{\vartheta}{2}}} = \frac{1}{3} \omega^2 \tilde{\rho}^3 \cos^3{\theta} + \cos{\theta} + \ln{\tan{\frac{\theta}{2}}}.
\label{eq:curly}
\end{equation}
ER11 tells us that the range of $F$ for a given $\omega$ is $[F_1, F_0]$, where
\begin{equation}
F_0 \equiv F\left(0\right) = \left[1-\omega^2\,\tilde{\rho}\left(0\right)^3\right]^{-2/3} = \left(1-\omega^2\right)^{-2/3}
\label{eq:F0}
\end{equation}
and
\begin{equation}
F_1 \equiv F\left(1\right) = e^{\frac{2}{3}\omega^2\,\tilde{\rho}\left(1\right)^3} = e^{\frac{2}{3}\omega^2 f^{-3}}
\label{eq:F1}
\end{equation}
with $F$ and $\tilde{\rho}$ seen as functions of $\tilde{z}$, $\tilde{\rho}\left(0\right) = 1$ and $\tilde{\rho}\left(1\right) = 1 / f$. Note that $F_0 \ge F_1 \ge 1$. In order to obtain $F$ from Equations \eqref{eq:F} and \eqref{eq:curly}, we define a new variable, $x\equiv\cos{\theta} \in [0, 1]$ and perform a change of variables from $\{\vartheta,\theta\}$ to $\{F, x\}$ in these equations. This results in
\begin{equation}
h(F;x,\omega) = 0,
\label{eq:nm}
\end{equation}
where 
\begin{equation}
h(F;x,\omega) \equiv \frac{x}{G} + \ln{\left(\frac{(1+x)\sqrt{F}}{x+G}\right)} - x - \frac{1}{3} \omega^2 \tilde{\rho}^3 x^3
\label{eq:h}
\end{equation}
with 
\begin{equation}
G \equiv G(F;x) \equiv \sqrt{x^2+F\left(1-x^2\right)}.
\label{eq:G}
\end{equation}
Thus, given $x$, we can treat $\tilde{\rho}$ as a function of $x$ and find $F$ by solving Equation \eqref{eq:nm} with $F$ as the independent variable. We do so using a variant of Newton's method, which requires
\begin{equation}
\frac{\partial h}{\partial F} = \frac{1}{2 F}\left(\frac{x}{G}\right)^3.
\label{eq:dhdF}
\end{equation}
We start half-way between $F_1$ and $F_0$ and add
\begin{equation}
\Delta F \left(F; x,\omega\right) = -\frac{h}{\partial h / \partial F} = 2F G^2\left[g_1 + g_2 + g_3\right]
\label{eq:dF}
\end{equation}
to our estimate of $F$ at each iteration, until $\Delta F$ is close to zero. Here,
\begin{equation}
g_1 \equiv g_1(F;x) \equiv \frac{G-1}{x^2},
\label{eq:g1}
\end{equation}
\begin{equation}
g_2 \equiv g_2(F;x) \equiv -\frac{G}{x^3}\ln{\frac{\left(1+x\right)\sqrt{F}}{x+G}},
\label{eq:g2}
\end{equation}
and
\begin{equation}
g_3 \equiv g_3(F;x,\omega) \equiv \frac{1}{3}\omega^2\tilde{\rho}^3G.
\label{eq:g3}
\end{equation}
A series expansion of each additive term in Equation \eqref{eq:dF} in $x$ around $x = 0$ shows that the equation's last term is $\mathcal{O}(1)$ and that its first two terms are
\begin{equation}
2F G^2 g_1 \approx 2\frac{F^{5/2}- F^2}{x^2}+\mathcal{O}\left(1\right)
\label{eq:g1term}
\end{equation}
and
\begin{equation}
2F G^2 g_2 \approx -2\frac{F^{5/2}- F^2}{x^2}+\mathcal{O}\left(1\right).
\label{eq:g2term}
\end{equation}
Consider what happens for $x \to 0$. As $F$ approaches the root of $h$ in the course of running the algorithm, $\Delta F \to 0$, so that the increasingly large terms in Equations \eqref{eq:g1} and \eqref{eq:g2} must cancel each other. Indeed, as implied by Equations \eqref{eq:dF}, \eqref{eq:g1term}, and \eqref{eq:g2term}, a series expansion of Equation \eqref{eq:dF} in $x$ around $x = 0$ doesn't have such terms:
\begin{multline}
\Delta F \approx \frac{2}{3} \left[F - F^{5/2} \left(1-\omega ^2\right)\right] + \\
x^2 \left[\frac{2}{5} - F^{3/2} \left(1-\omega ^2\right)+F^{5/2}\frac{3-8 \omega ^2}{5 \left(1-\omega ^2\right)}\right] + \mathcal{O}(x^4).
\label{eq:dFa}
\end{multline}
The coefficient of the $x^4$ term in this expansion is
\begin{multline}
\alpha_4(F; \omega) = \frac{1}{140} \Biggl[ 56-\frac{16}{F}-35 \sqrt{F} \left(1-\omega^2\right)-\\
14\,F^{3/2}\,\frac{1+4\,\omega ^2+10\,\omega ^4}{1-\omega ^2}+\\
F^{5/2}\,\frac{9 +8\,\omega ^2+44\,\omega ^4 \left(3-\omega ^2\right) }{\left(1-\omega ^2\right)^3} \Biggr].
\label{eq:alpha}
\end{multline}
Under constant machine epsilon $q$, decreasing $x$ increases the absolute error in $\Delta F$ (and thus in $F$) associated with the increasingly large additive terms in Equation \eqref{eq:dF}. At the same time, the error associated with Equation \eqref{eq:dFa} decreases, since the series expansion becomes a better approximation of $\Delta F$. We seek to approximate $\Delta F$ using the better of Equations \eqref{eq:dF} and \eqref{eq:dFa}, setting the boundary between the methods at the point where their respective errors are equal. 

Given Equations \eqref{eq:g1term} and \eqref{eq:g2term}, the rounding error due to Equation \eqref{eq:dF} is
\begin{equation}
\varepsilon_{A} = 2 \frac{q}{A} \frac{F^{5/2}- F^2}{x^2}.
\label{eq:erFull}
\end{equation}
where $A \ge 1$ and $q$ is machine epsilon, $\sim$2$\times$10$^{-16}$ in double precision.  We also approximate the error associated with using Equation \eqref{eq:dFa} as
\begin{equation}
\varepsilon_{B} = B \alpha_4 x^4,
\label{eq:erApprox}
\end{equation}
where $B \sim 1$. We then equate Equations \eqref{eq:erFull} and \eqref{eq:erApprox}, approximate $F$ with $F_0$, define $k \equiv A B$, and solve for $x$:
\begin{equation}
x = \left(\frac{2q}{k} \, \frac{F_0^{5/2}- F_0^2}{\alpha_4(F_0)}\right)^{1/6}.
\label{eq:xb}
\end{equation}
We cast the right-hand side of Equation \eqref{eq:xb} as a function of $\omega$ with the help of Equations \eqref{eq:F0} and \eqref{eq:alpha}. As $\omega$ approaches 0, the computation of this expression becomes impossible due to rounding error. Accordingly, we sum the first three terms of its series expansion around $\omega = 0$ to obtain $x_b$, a boundary value of $x$:

\begin{multline}
x_b(\omega) = q^{1/6} \left(\frac{2}{6885 k}\right)^{1/6} \Biggl[\,3\, \omega^{-2/3} -\\ \frac{199}{255}\,\omega^{4/3} - \frac{29123}{65025}\,\omega^{10/3}\,\Biggr].
\label{eq:xbApprox}
\end{multline}
As $\omega$ decreases even further, Equation \eqref{eq:xbApprox} exceeds 1, the upper bound of $x$'s range. For these values of $\omega$, we set $x_b$ to 1.

\begin{figure}
\includegraphics[width=\linewidth]{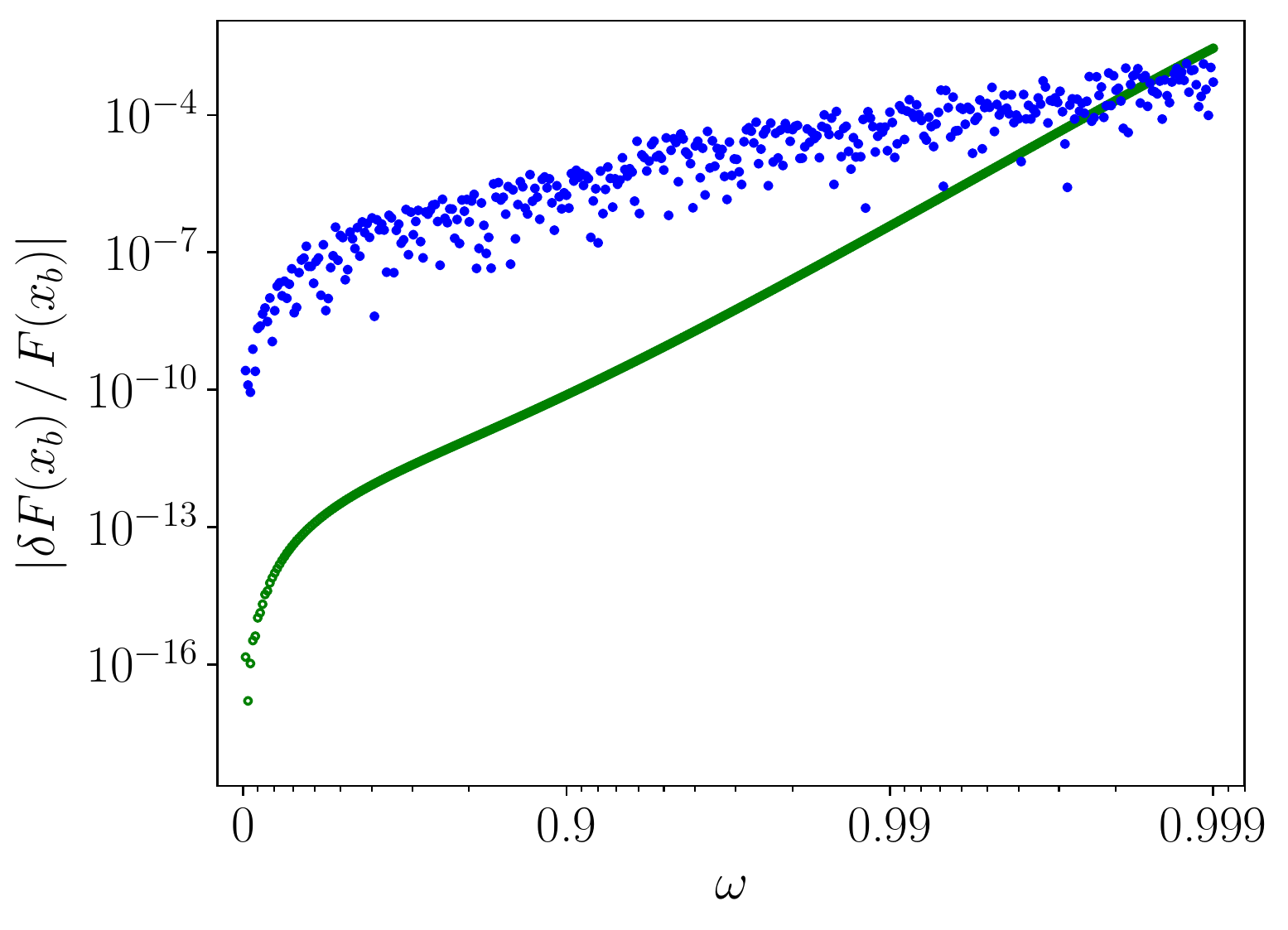}
\caption{Relative error in $F$ at $x_b$ for floating point resolution $q = 2\times10^{-16}$, $\omega \in [0, 0.999]$ and parameter $k = 100$ (see Section \ref{subsec:tgcalc}). Blue filled and green open markers correspond to the use of Equation \eqref{eq:dF} and \eqref{eq:dFa}, respectively.}
\label{fig:erF}
\end{figure}

To estimate the actual error in $F$, we use equation \eqref{eq:dF} to calculate its etalon values with \texttt{mpmath} \citep{mpmath} and $q \sim 10^{-100}$. In the rest of our calculations, we use equation \eqref{eq:dFa} up to the second order in $x$ whenever $x \le x_b$ and equation \eqref{eq:dF} otherwise, with $q = 2\times10^{-16}$ throughout. If $\omega\in [0, 0.999]$, the resulting relative error in $F$ at $x_b$ tends to its maximum value at $\omega = 0.999$. We set $k$ to $100$, where this maximum error is close to minimized, equal to $0.3\%$  (see Figure \ref{fig:erF}). According to equation \eqref{eq:T}, this corresponds to a lower relative error in $T$.

For large enough values of $\omega$ in its range, an iteration of both an error-free Newton's method and our error-prone algorithm can result in $F < F_1$. In such cases, we set $F$ to $F_1$. Additionally, we set $F$ to $F_0$ if $F > F_0$ in the course of running the error-prone algorithm, which can happen at low $\omega$.

The algorithm converges within approximately 5, 6, 8, and 10 iterations for $\omega \le 0.9$, $\omega = 0.95$, $\omega = 0.99$, and $\omega = 0.999$, respectively. In this article's work, we run it with 15 iterations, which takes no more than 0.1\% of the total computing time for a 1221-wavelength spectrum.

\subsection{Coefficient interpolation} \label{subsec:interp}

We now define the temperature-dependent Planck factor as 
\begin{equation}
    {\cal P}(T) \equiv \left( \exp{\left[\frac{h c}{\lambda k_B T}\right]} - 1 \right)^{-1},
    \label{eq:planckfactor}
\end{equation}
where $h$ is Planck's constant, $c$ is the speed of light in vacuum, and $k_B$ is Boltzmann's constant.

In Section \ref{subsec:fits} we obtain coefficients $a_{ij}$ on a discrete grid of $g$ and $T$. However, the values of $g$ and $T$ we get via Equations \eqref{eq:g} and \eqref{eq:T} are not necessarily on that grid. To obtain the $a_{ij}$ that enter Equation \eqref{eq:int3}, we interpolate each coefficient linearly in either $g$ or $\log{g}$ and in either $T$, $\log{T}$, or ${\cal P}(T)$.

\begin{figure*}
\includegraphics[width=0.5\linewidth]{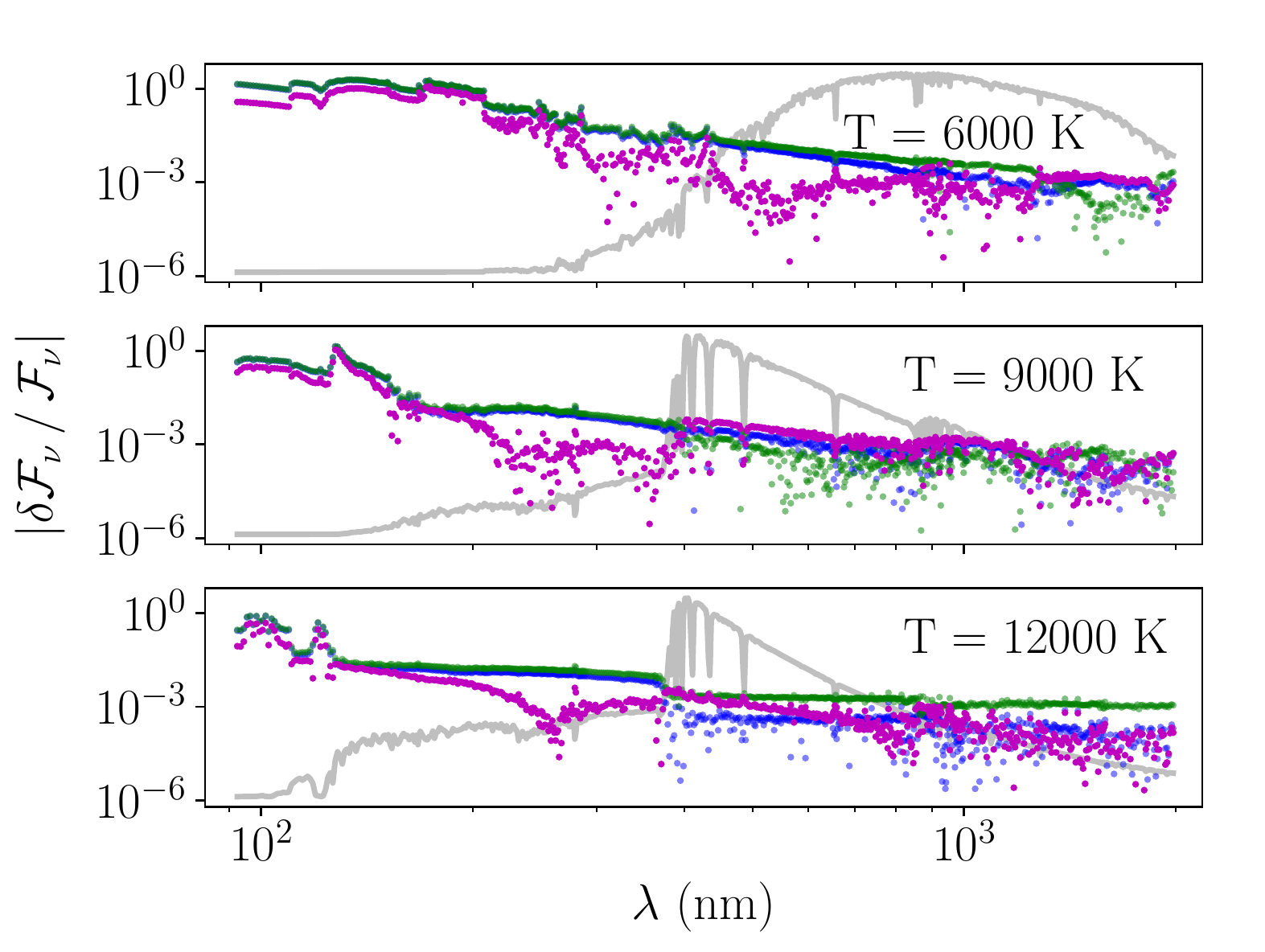}
\includegraphics[width=0.5\linewidth]{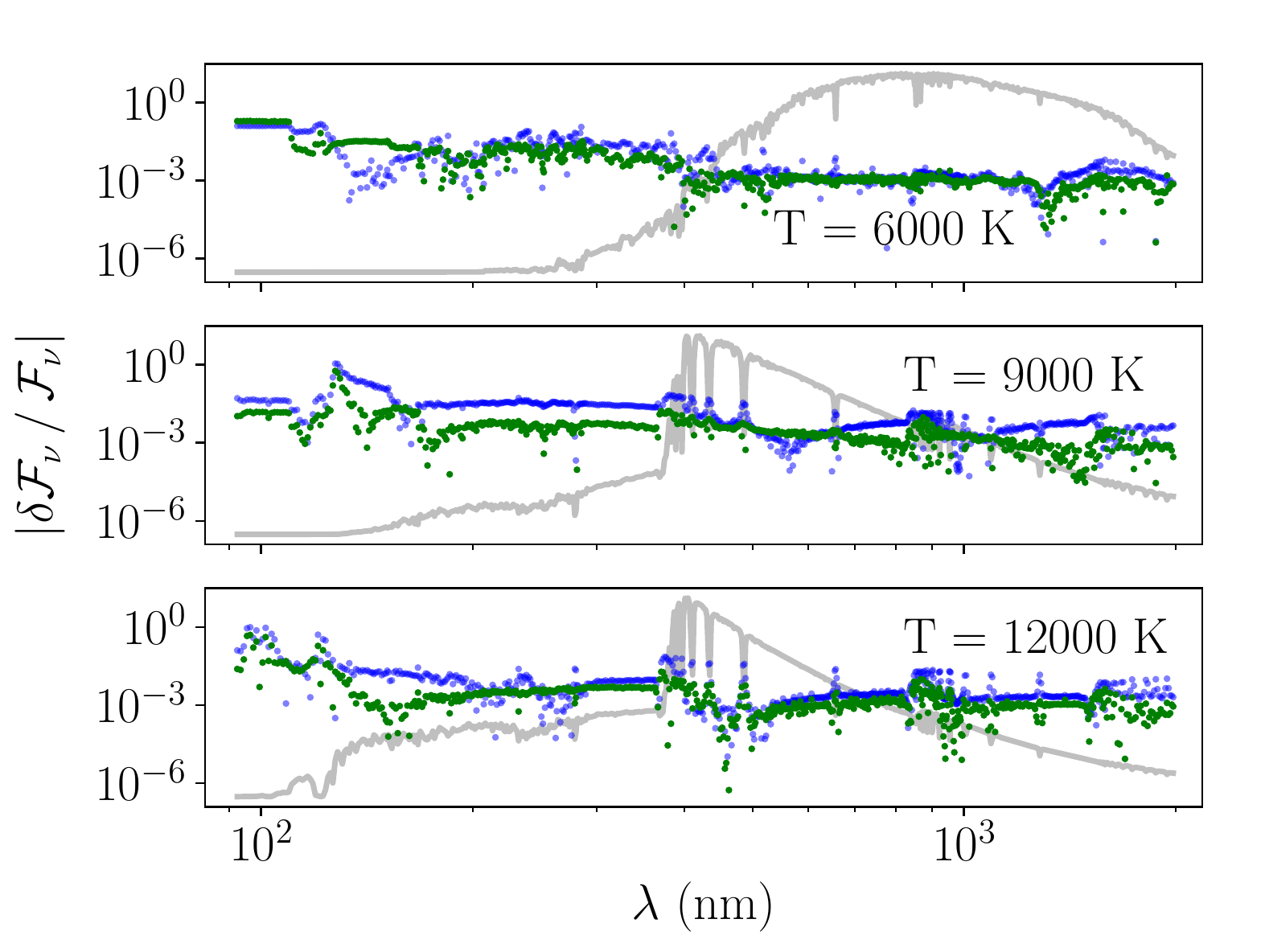}
\caption{Relative flux errors due to interpolation in temperature (left panels) and gravity (right panels) for three nonrotating stars with $\log_{10} g = 4.5$ (cgs) and $T = 6000$, $9000$, and $12000$~K. Blue markers indicate interpolation in $T$ and $g$, green --- in $\log{T}$ and $\log{g}$, magenta --- in ${\cal P}(T)$ (see Equation \eqref{eq:planckfactor}). Grey lines show the stellar spectra on linear (not logarithmic) vertical scales. 
Errors are typically $\lesssim$1\%, often much lower, in the spectral regions responsible for a significant fraction of the stellar flux. We compute these errors by omitting and interpolating over a tabulated model, artificially making them twice as far away from grid points as they would otherwise be. Real interpolation errors will be substantially lower.}
\label{fig:erT}
\end{figure*}

In order to assess the accuracy of interpolation, we aim to compute fractional errors in intensity for three fiducial nonrotating stars with $T = 6000$, 9000, and 12000~K, all with $g = 10^{4.5} \mathrm{cm/s^2}$. Our choices of $g$ and $T$ are on the grid in CK04, so that we can first compute the stars' spectra without interpolation error. We also compute the stars' spectra with intensity information missing, either at their values of $T$ or at their values of $g$, interpolating between the closest neighboring values where such information is available. The neighboring temperature pairs are $\{5750, 6250\}$, $\{8750, 9250\}$, and $\{11500, 12500\}$ K for the above-listed three stars, respectively. The neighboring gravity values are $10^{4.0}$ and $10^{5.0}$ $\mathrm{cm/s^2}$.
Figure \ref{fig:erT} demonstrates the relative error we infer from the comparison of error-free spectra and spectra that involve interpolation. For the most part, ${\cal P}(T)$ and $\log{g}$ are the best interpolants, although others can be better at some $\lambda$ and $T$. The $\log g$ and $T$ differences between known points in the above accuracy assessments are twice their normal values, so that our assessments over-estimate interpolation error (by a factor $\sim$4 assuming that linear and quadratic terms dominate the local series expansions). 

\section{Longitudinal integral} \label{sec:integr}

\subsection{Numerical schemes} \label{subsec:numerical}

In order to approximate the integral in Equation \eqref{eq:int3}, we first evaluate its integrand, 
\begin{equation}
f(\tilde{z}) \equiv 2 R_{eq}^2 \, \tilde{A}(\tilde{z}) \sum_{j,k} a_{jk}(\tilde{z})\, P_{jk}(\tilde{z}),
\label{eq:integrand}
\end{equation}
at a set of ${\cal N} + {\cal N}_l - 1$ equally spaced discrete values $\{\tilde{z}_i\}$. Here, $i \in \{-{\cal N}_l + 1,\, -{\cal N}_l + 2,\, \ldots,\, 0, \,\ldots,\, {\cal N} - 2,\, {\cal N} - 1\}$, $\tilde{z}_{{\cal N} - 1} = 1$, and $\tilde{z}_i < \tilde{z}_j$ when $i < j$. We ensure that $\tilde{z}_0 = 0$ and define $\Delta\tilde{z} \equiv \tilde{z}_1 - \tilde{z}_0$. Furthermore, we define $\delta\tilde{z} \equiv \tilde{z}_{-{\cal N}_l + 1} - (- \tilde{z}_b)$ and note that ${\cal N}_l$ is a function of ${\cal N}$, as it satisfies $-\tilde{z_b} \le \tilde{z}_{-{\cal N}_l + 1} < -\tilde{z_b} + \Delta\tilde{z}$. The evaluation of $f_i \equiv f(\tilde{z}_i)$ for a given $i$ is described in Sections \ref{sec:surface} and \ref{sec:intensity}. It is possible that $-\tilde{z}_b \ne \tilde{z}_i$ for all $i$, so that we do not evaluate $f(-\tilde{z}_b)$ directly. At the same time, $\mu = 0$ at $\tilde{z} = -\tilde{z}_b$, so that $P_{ij}(-\tilde{z}_b) = 0$ for all $\{i,j\}$ (see equations \eqref{eq:p} and \eqref{eq:P}). Thus, $f(-\tilde{z}_b) = 0$. 

We wish to construct a numerical integration scheme that approximates $f(\tilde{z})$ as a piecewise polynomial of up to third order. If we had an analytic expression for the integrand, we might have been able to predict the performance of a possible scheme by evaluating the integrand's derivatives. Since we do not have such an expression, we start by examining the integrand at many discrete values of $\tilde{z}$ for a number of different stars. The integrand looks amenable to numerical integration almost everywhere. The exception is a downward cusp at $\tilde{z} = 0$ when $\omega \to 1$. This cusp is partly due to the fact that
\begin{equation}
\lim_{\substack{\omega \to 1 \\ \tilde{z}\to0^\pm}} \tilde{r}'(\tilde{z}) = \mp \frac{1}{f \sqrt{3}},
\label{eq:limit}
\end{equation}
which results from equations \eqref{eq:Ds} and \eqref{eq:rprime}, evaluated at the given limits. Here, $f$ is a constant, defined in Section \ref{subsec:shape}. This discontinuity in $\tilde{r}'$ leads to discontinuities in both $\{P_{ij}\}$ and $\tilde{A}$ (see equations \eqref{eq:P}, \eqref{eq:p}, \eqref{eq:area3}, \eqref{eq:n}, and \eqref{eq:mu}), and thus contributes to the cusp in $f(\tilde{z})$. There is another contributing factor. When $\omega \to 1$, both $T$ and $g$ approach zero at the equator and increase quickly away from zero latitude (see equations \eqref{eq:g} and \eqref{eq:T}). This causes $\{a_{ij}\}$ to also change rapidly with latitude in the equatorial regions under these conditions (see Section \ref{subsec:fits}).

Thus, when $\omega$ approaches 1, the shape of the star and its effective gravity profile near $\tilde{z} = 0$ jointly lead to a cusp in the integrand. In view of this fact, we split the integral into two parts. We use $\{f_i\}$ with $i \in \{-{\cal N}_l + 1,\, \ldots,\, 0\}$ and with $i \in \{0,\, \ldots,\, {\cal N} - 1\}$ to approximate the integral on the lower interval $[-\tilde{z}_b, 0]$ and on the upper interval $[0, 1]$, respectively. To calculate the integral on the upper interval ${\cal I}_u$, we make use of an approximation based on the fitting of cubic polynomials through successive groups of four points \citep[equation 4.1.14 in][]{recipes_2007}:
\begin{multline}
{\cal I}_u \equiv \int^1_0 f(\tilde{z})\,d\tilde{z} = \Delta\tilde{z}\, \Biggl[\frac{3}{8}f_0+\frac{7}{6}f_1+\frac{23}{24}f_2+f_3+\\
\ldots+f_{{\cal N}-4}+\frac{23}{24}f_{{\cal N}-3}+\frac{7}{6}f_{{\cal N}-2}+\frac{3}{8}f_{{\cal N}-1}\Biggr] + \varepsilon_u({\cal N}),
\label{eq:intup}
\end{multline}
where $\varepsilon_u({\cal N})$ is the upper-interval error term. 

When ${\cal N}_l \ge 2$ and $\delta\tilde{z} \ne 0$, we calculate the quadratic polynomial that goes through points $\{(-\tilde{z}_b, 0), (\tilde{z}_{-{\cal N}_l + 1}, f_{-{\cal N}_l + 1}), (\tilde{z}_{-{\cal N}_l + 2}, f_{-{\cal N}_l + 2})\}$ and shift it horizontally to go through $(0,0)$, so that its form becomes $a\tilde{z}^2 + b\tilde{z}$. 

When both the inclination and ${\cal N}$ are small, it is possible that ${\cal N}_l = 1 \vee \left({\cal N}_l = 2 \wedge \delta\tilde{z} = 0\right)$. In this case, we approximate the integrand by a straight line between $(-\tilde{z}_b, 0)$ and $(0, f_0)$, so that the integral on the lower interval ${\cal I}_l$ is
\begin{equation}
{\cal I}_l \equiv \int^0_{-\tilde{z}_b} f(\tilde{z})\,d\tilde{z} = \frac{1}{2}\,\delta \tilde{z}\,f_0  + \varepsilon_l({\cal N}),
\label{eq:intdn1}
\end{equation}
where $\varepsilon_l({\cal N})$ is the lower-interval error term. 

When ${\cal N}_l = 2$ and $\delta\tilde{z} \ne 0$, we approximate ${\cal I}_l$ by the integral of the above-mentioned quadratic on $[0, \tilde{z}_b]$:
\begin{equation}
{\cal I}_l = \frac{1}{3}\,a\, \tilde{z}_b^3  + \frac{1}{2}\,b\, \tilde{z}_b^2 + \varepsilon_l({\cal N}).
\label{eq:intdn2}
\end{equation}
When ${\cal N}_l \ge 3$, we use the quadratic to approximate the integral up to the lowest $\tilde{z}_i$:
\begin{equation}
\int^{\tilde{z}_{-{\cal N}_l + 1}}_{-\tilde{z}_b} f(\tilde{z})\,d\tilde{z} \approx \delta{\cal I} \equiv \frac{1}{3}\,a\, (\delta\tilde{z})^3  + \frac{1}{2}\,b\, (\delta\tilde{z})^2.
\label{eq:intdelta}
\end{equation}
For ${\cal N}_l$ = 3, 4 and 5, we approximate the rest of ${\cal I}_l$ by Simpson's rule, Simpson's 3/8 rule and Boole's rule, respectively \citep[see Section 4.1.1 of][]{recipes_2007}. For example, when ${\cal N}_l = 5$,
\begin{multline}
{\cal I}_l = \frac{\Delta\tilde{z}}{45}\, \Biggl[14 f_{-4}+64 f_{-3}+\\24 f_{-2}+ 64 f_{-1}+ 14 f_0\Biggr] + \delta{\cal I} + \varepsilon_l({\cal N}).
\label{eq:intdn5}
\end{multline}
When ${\cal N}_l \ge 6$, we combine equation \eqref{eq:intdelta} with the approximation in equation \eqref{eq:intup}:
\begin{multline}
{\cal I}_l = \Delta\tilde{z}\times\\
\Biggl[\frac{3}{8}f_{-{\cal N}_l+1}+\frac{7}{6}f_{-{\cal N}_l+2}+\frac{23}{24}f_{-{\cal N}_l+3}+f_{-{\cal N}_l+4}+\\
\ldots+f_{-3}+\frac{23}{24}f_{-2}+\frac{7}{6}f_{-1}+\frac{3}{8}f_{0}\Biggr] + \delta{\cal I} + \varepsilon_l({\cal N}).
\label{eq:intdn6}
\end{multline}
The error terms, $\varepsilon_u({\cal N})$ and $\varepsilon_l({\cal N})$, can both be seen as functions of ${\cal N}$ for a given star. The integration scheme corresponding to equations \eqref{eq:intup}--\eqref{eq:intdn6} involves approximating the integrand by cubic polynomials everywhere except for the small interval in equation \eqref{eq:intdelta} and the entire lower interval when ${\cal N}_l \le 3$. In the foregoing, we designate this scheme as cubic. 

An alternate integration scheme is an application of the trapezoidal rule to separately calculate ${\cal I}_u$ and ${\cal I}_l$, whilst setting the integral in equation \eqref{eq:intdelta} to zero. We designate this scheme as trapezoidal. In either scheme, the flux is calculated according to 
\begin{equation}
D_{\star}^2\,\mathcal{F}_\nu = {\cal I}_u + {\cal I}_l,
\label{eq:flux}
\end{equation}
where the unknown error terms on the right-hand side are set to zero.

\subsection{Convergence} \label{subsec:convergence}

\begin{figure}
\includegraphics[width=\linewidth]{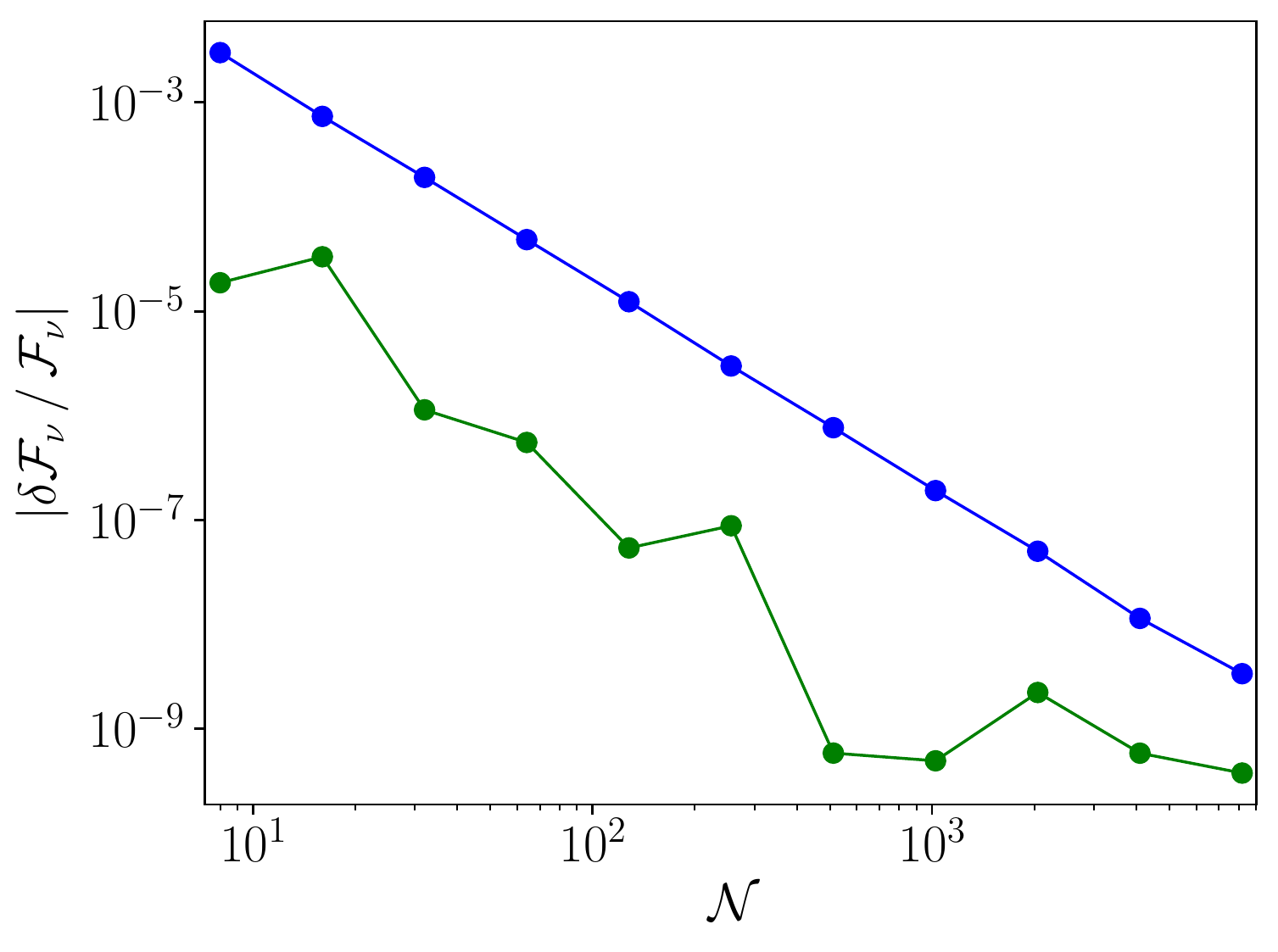}
\caption{Relative flux error due to integration in $\tilde{z}$ for synthetic Vega at $i = \pi / 4$ and $\lambda = 511$ nm as a function of the number of abscissae at the equator and above (see Section \ref{sec:integr}). Blue and green points correspond to the trapezoidal and cubic approximations, respectively.}
\label{fig:erz}
\end{figure}

\begin{figure*}
\includegraphics[width=0.45\linewidth]{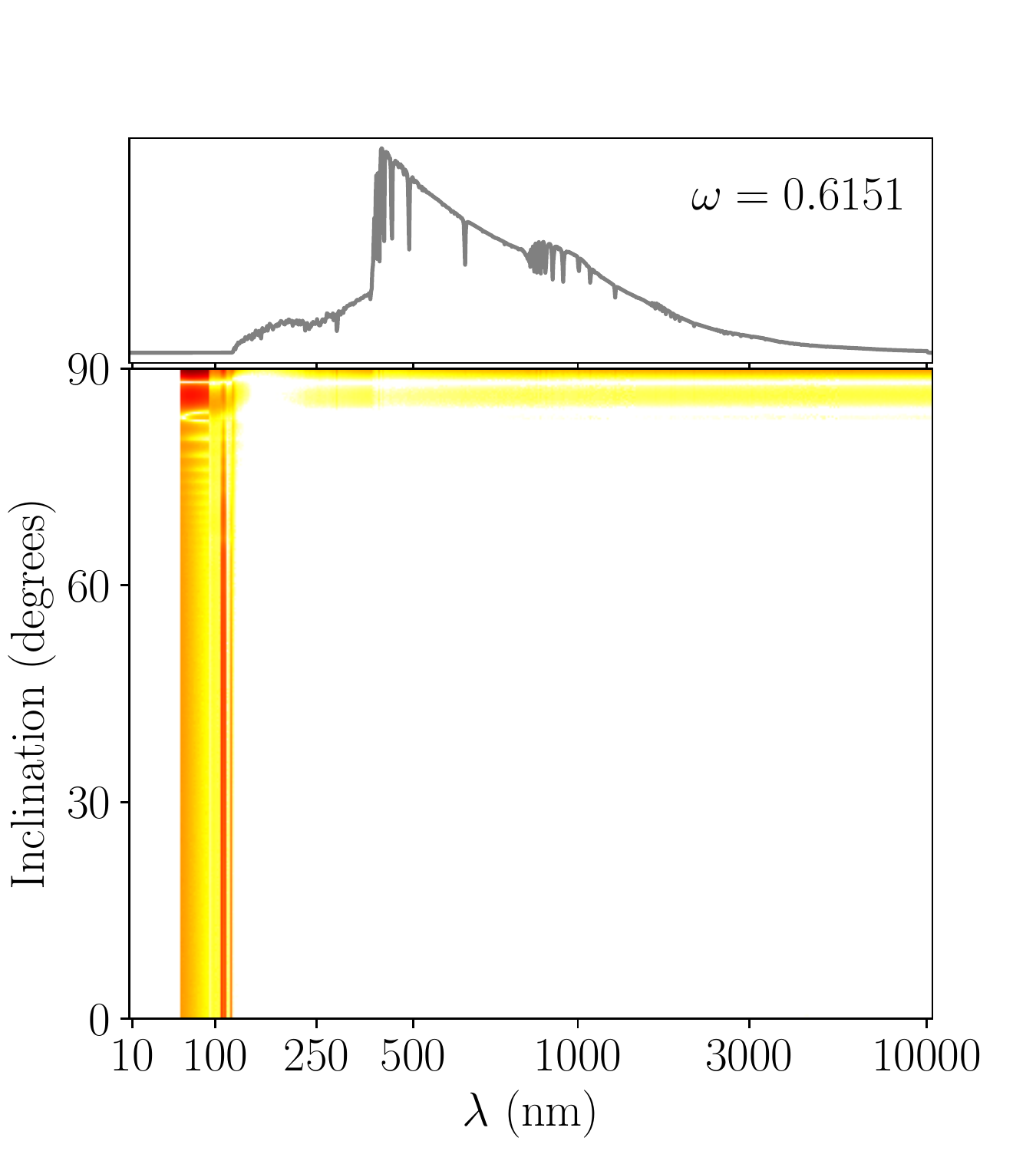}
\includegraphics[width=0.45\linewidth]{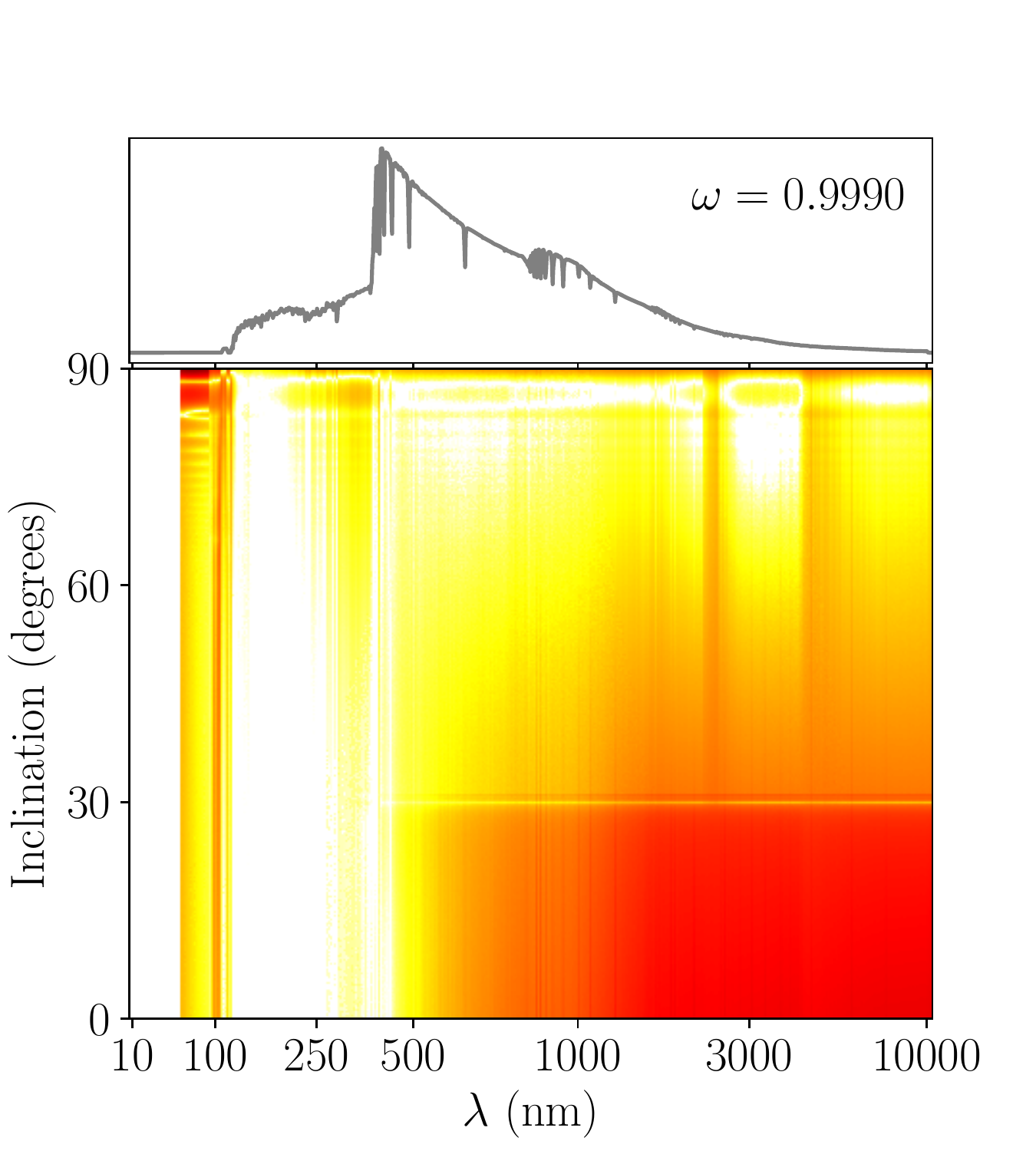}
\includegraphics[width=0.09\linewidth]{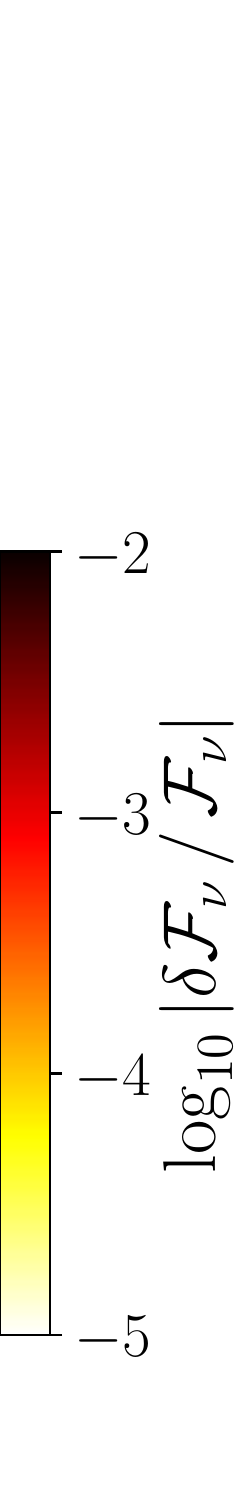}
\caption{Bottom panels: relative flux error due to the cubic approximation of the integral in $\tilde{z}$ with ${\cal N} = 100$  for two synthetic stars with Vega's $M$, $L$, and $R_e$. See Section \ref{sec:integr} for details. Top panels: the stars' spectra at $i = 0\degree$ (pole-on). The star on the left has Vega's $\omega$ from YP10. The star on the right has $\omega = 0.999$; below 100 nm, its spectra do not rise above 0.02\% of their maxima; at 100 nm and above, the maximum error from our discretization in $\tilde{z}$ is 0.1\%.}
\label{fig:erzall}
\end{figure*}

To assess the convergence properties of the two schemes, we synthesize a star with Vega's $M$, $L$, $R_{e}$ and $\omega$ from YP10. When the synthetic star is seen at $i = \pi / 4$, $\lambda = 511$ nm is close to its spectral peak. We compute ${\cal F}_\nu({\cal N})$, the flux for this combination of $i$ and $\lambda$ at different ${\cal N}$ and plot
\begin{equation}
\left|\frac{\delta{\cal F}_\nu}{{\cal F}_\nu}\right| \equiv \frac{\left|{\cal F}_\nu({\cal N}) - {\cal F}_\nu(10,000)\right|}{{\cal F}_\nu(10,000)}
\label{eq:fluxerr}
\end{equation}
in Figure \ref{fig:erz}. The trapezoidal approximation converges according to a power law, the cubic approximation does better, and the two approximations converge at about the same rate. The error due to either approximation is well below 0.1\% when ${\cal N} \ge 100$. We  present the error due to the cubic approximation at ${\cal N} = 100$ for the remaining wavelengths and inclinations in the lower left panel of Figure \ref{fig:erzall}. The picture in Figure \ref{fig:erz} is typical across wavelengths and stars, although the advantage of the cubic approximation and the performance of each scheme all tend to decrease as $\omega \to 1$. The latter behavior is not surprising, given the discussion of the associated limit earlier in this Section. To characterize a possibly worst-case scenario, we compute the error in equation \eqref{eq:fluxerr} for a star that combines $\omega = 0.999$ with Vega's $M$, $L$, and $R_{e}$. The lower right panel of Figure \ref{fig:erzall} present the result of this calculation on a grid of inclinations, for each wavelength in CK04. The error remains no higher than 0.1\% for all wavelengths above 100 nm, where the spectra are appreciably nonzero. We use ${\cal N} = 100$ for all models in this work, unless stated otherwise.

\begin{figure*}
\includegraphics[width=0.5\linewidth]{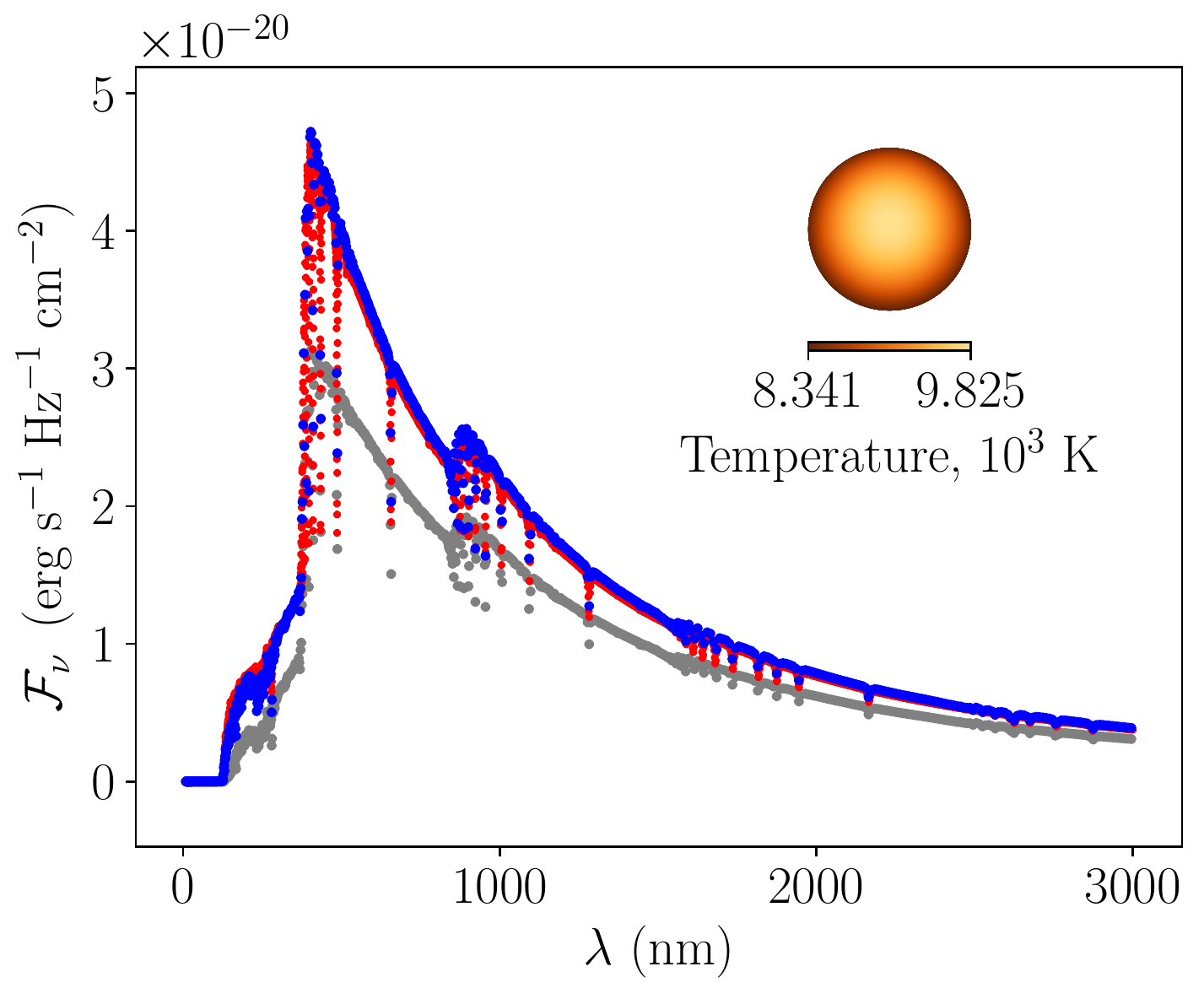}
\includegraphics[width=0.5\linewidth]{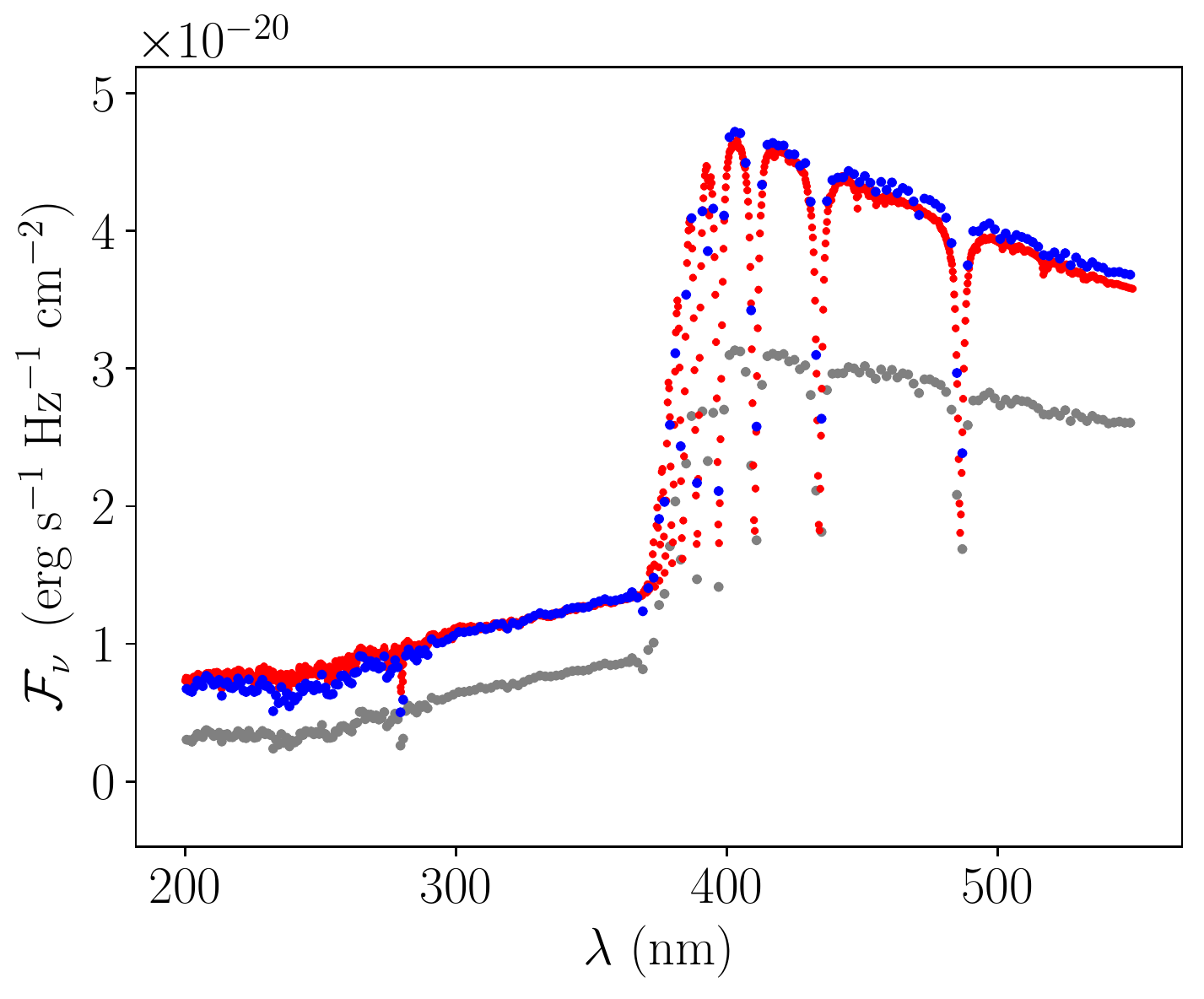}
\caption{Left panel: Synthetic spectrum of Vega computed using its observed parameters (blue points) compared to the observed spectrum (red points). The two spectra agree well. The gray points correspond to a synthetic Vega viewed edge-on and show a significantly fainter and redder star. The inset shows the local effective temperature across the visible surface of the synthetic star at its observed inclination of $\sim$5$^\circ$ (YP10). Right panel: same as left panel, with wavelength range restricted to the Balmer jump.}
\label{fig:spectra}
\end{figure*}

Figure \ref{fig:spectra} compares the observed spectrum of Vega from \citet{bohlin_2014} with the star's synthetic spectrum at the inclination from YP10 and distance from \citet{hipparcos_2007}. It also shows the synthetic star's spectrum at $i = \pi/2$ (were we to view Vega edge-on), with lower intensity and a redder spectrum indicative of the cooler equatorial regions. The observed spectrum and the synthetic spectrum at YP10's inclination are quite close, indicating the accuracy of the synthetic star's parameters. Figure \ref{fig:spectra} may be compared with Figures 8 and 9 in \citet{aufdenberg_2006_apj}, where the parameters of a stellar model are fit to Vega's observed spectrum.

Given pre-computed $\{a_{kj}\}$ on CK04's parameter grid in Section \ref{subsec:fits}, we split the remaining computation for a specific star into an inclination-independent and an inclination-dependent portion. The former includes the calculation of $\tilde{r}(\tilde{z})$ and $\tilde{A}(\tilde{z})$ in Section \ref{subsec:shape}, $g(\tilde{z})$ and $T(\tilde{z})$ in Section \ref{subsec:tgcalc}, and $\{a_{kj}(\tilde{z})\}$ in Section \ref{subsec:interp} (via interpolation over $T$ and $g$). The inclination-dependent portion includes the calculation of $\tilde{z}_b$ in Section \ref{subsec:vis}, $\{P_{kj}(\tilde{z})\}$ in Section \ref{subsec:piecewise}, and the one-dimensional integral in this section. On a 2.3 GHz MacBook Pro with 8 GB of RAM, inclination-independent computation with ${\cal N} = 100$ takes about 800 ms. Thereafter, the inclination-dependent computation takes about 30 ms per inclination, so that the full synthetic spectrum at 50 inclinations takes $\sim$2~seconds.

\section{Extensions} \label{sec:ext}

\subsection{Color-magnitude diagrams} \label{subsec:colormag}

Photometry is available for many more stars, and star clusters, than spectroscopy. As a result, much of the recent work studying the observational consequences of rapid stellar rotation has used color-magnitude diagrams \citep[e.g.][]{Bastian+deMink_2009,brandt_2015_apj25,DAntona+DiCriscienzo+Decressin+etal_2015,Goudfrooij+Girardi+Correnti_2017,gossage_2019_apj}. Our spectra may be easily used to compute colors and magnitudes in any photometric system.

Here, we briefly show the consequences of inclination for rapidly rotating stars in the color-magnitude diagram. For a given filter in \citet{rodrigo_2012}, we approximate the transmission curve $T(\lambda)$ via third-order spline interpolation between the discrete points with available transmission values. We also convert the previously calculated ${\cal F}_\nu(\lambda)$ to $\mathcal{F}_{\lambda}(\lambda)$:
\begin{equation}
\mathcal{F}_{\lambda}(\lambda) = \frac{c}{\lambda^2}\, {\cal F}_\nu(\lambda).
\label{eq:conv}
\end{equation}
Next, we estimate the star's flux through the filter,
\begin{equation}
{\cal F} = \int_0^{\infty} \mathcal{F}_{\lambda}(\lambda)\, T(\lambda) \, d\lambda,
\label{eq:filt}
\end{equation}
via the application of the trapezoidal rule over the variable-size intervals between nearest-neighbor $\lambda$ values in CK04. Finally, we calculate the magnitude:
\begin{equation}
m = -2.5 \log_{10}{\frac{{\cal F}}{{\cal F}_0\int_0^{\infty} T(\lambda) \, d\lambda}},
\label{eq:mag}
\end{equation}
where ${\cal F}_0$ is the flux zero point in the Vega calibration system and the integral is approximated the same way as the one in equation \eqref{eq:filt}.

\begin{figure}
\includegraphics[width=\linewidth]{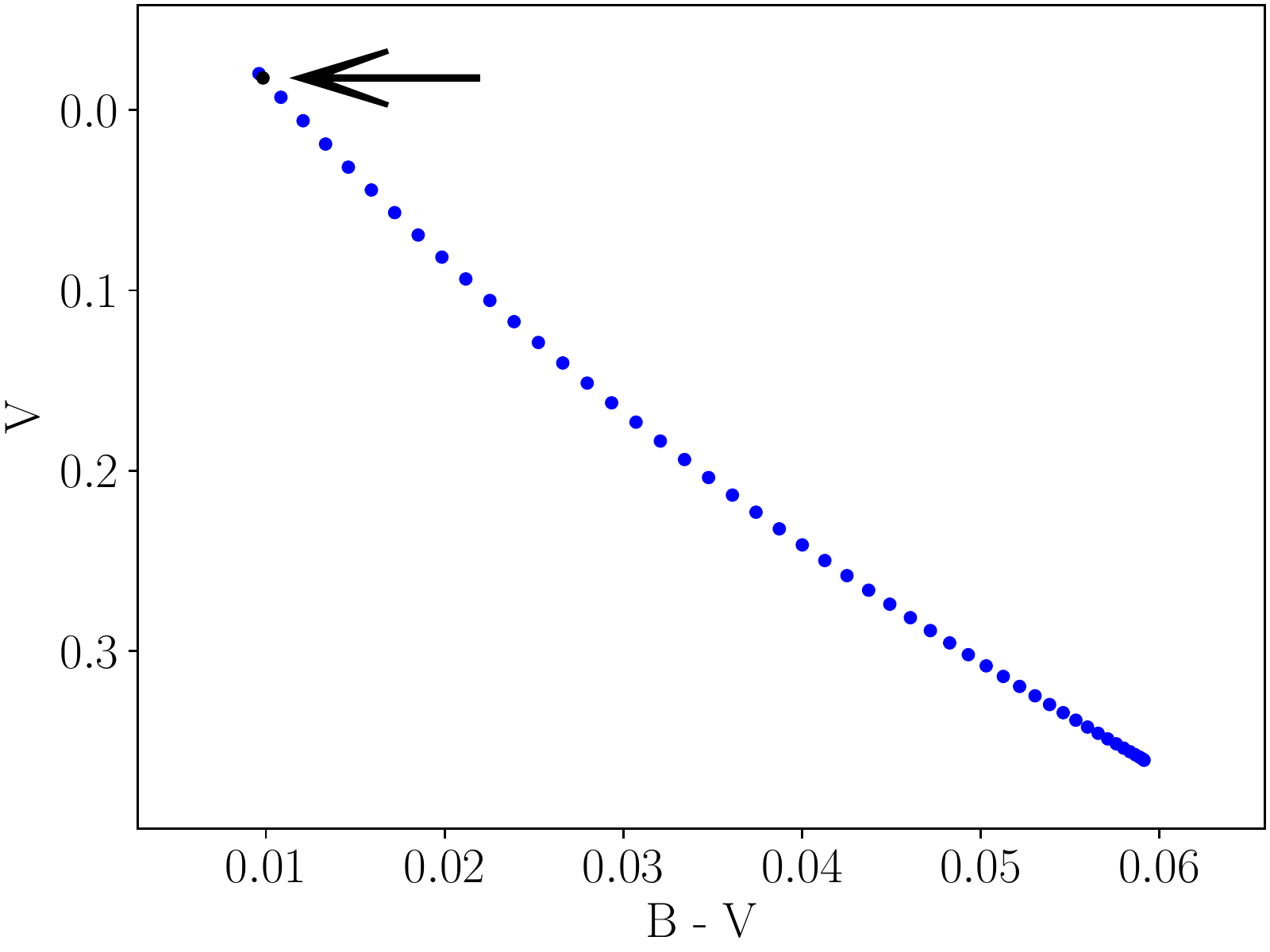}
\caption{Visual magnitude versus the difference between blue and visual magnitudes of synthetic Vega. From left to right, inclination increases from 0 to $\pi / 2$. Markers are spaced evenly in $\cos{i}$, corresponding to isotropically distributed $i$ \citep[e.g.,][]{corsaro2017natas}. The black marker and arrow indicate the star's observed inclination.}
\label{fig:colormag}
\end{figure}

We calculate the magnitudes corresponding to the Generic Bessel $B$ and $V$ filters in \citet{rodrigo_2012} for a single star with Vega's physical parameters, but observed at inclinations from 0 to $\pi/2$. Figure \ref{fig:colormag} shows the resulting range of $(B - V, V)$. The black arrow indicates Vega's actual inclination. A similar calculation with a range of stellar models would produce a smooth distribution suitable for comparison with observed color-magnitude diagrams.

\subsection{Planetary transits} \label{subsec:transit}

\subsubsection{Introduction}

Rapid stellar rotation can have an observable effect on the light curve of a planetary transit: the transit will be deeper or shallower as the planet transits hotter or cooler parts of the stellar surface \citep{barnes_2009}. This effect can constrain the projected obliquity, the angle between the star's angular momentum, and the planet's orbital angular momentum \citep{barnes_2009,Barnes+Linscott+Shporer_2011,Masuda_2015}. In this section we extend our tool to rigorously compute the light curve of a planet obliquely transiting a rapidly rotating star.

Consider a planet of radius $R_1$ that transits its host star. In our coordinate system, $\mathbf{\hat{y}} \perp \mathbf{\hat{i}}$. We let $\mathbf{\hat{z}'} \equiv \mathbf{\hat{i} \times \hat{y}}$ and refer to the plane spanned by $\mathbf{\hat{y}}$ and $\mathbf{\hat{z}'}$ as the view plane. Over the course of the transit and in projection onto the view plane, the planet's center traces a straight line with projected impact parameter $b$ with respect to the star's center and projected obliquity $\alpha \in [-\pi/2, \pi/2]$ with respect to the star's rotational axis.

\subsubsection{Intensity at a sight line}

Slices of the stellar surface perpendicular to the z-axis are circles with centers $(0, 0, z)$ and radii $r$, with $z \in [-R_p, R_p]$. Their projections onto the view plane are ellipses that satisfy
\begin{equation}
\left( \frac{z' - z \sin{i}}{r \cos{i}} \right)^2 + \left( \frac{y}{r} \right)^2 = 1.
\label{eq:ellipse1}
\end{equation}
With $u' \equiv z' / R_e$ and $\tilde{y} \equiv y / R_e$, Equation \eqref{eq:ellipse1} can be re-written as
\begin{equation}
\tilde{r}^2 = \left(u'\,\sec{i} - u\, \tan{i}\right)^2 + \tilde{y}^2.
\label{eq:ellipse2}
\end{equation}
Given a sight line through point $(\tilde{y}, u')$ in the view plane, we wish to find the point $(\tilde{r}, \phi, u)$ on the stellar surface where the sight line's light originates. Pairs of $u \in [-z_b / R_e, z_b / R_e]$ and $\tilde{r} \in [0, 1]$ that satisfy both Equation \eqref{eq:surf} and Equation \eqref{eq:ellipse2} correspond to the surface points that the sight line pierces. To find these points, we substitute the right-hand side of Equation \eqref{eq:ellipse2} for $\tilde{r}^2$ in Equation \eqref{eq:surf} and perform algebraic manipulation that produces a 6th degree polynomial equation in $u$. We then substitute this equation's real roots for $u$ in Equation \eqref{eq:ellipse2} and pick out the ones that correspond to both $u$ and $\tilde{r}^2$ in their respective ranges. There are two roots in the latter category when the sight line pierces the star and zero such roots when it doesn't. In the former case, the greater root gives the $u$ and, via Equation \eqref{eq:ellipse2}, the corresponding $\tilde{r}$ coordinates for the sight line.

Only one additive term, 
\begin{equation}
\frac{(u' \, \sec{i})^6\, \omega^4}{4},
\label{eq:sec}
\end{equation}
contains the highest power of $\sec{i}$ in the 6th degree polynomial. By a heuristic analogue of the argument in Section \ref{subsec:tgcalc}, computation of $u$ becomes impossible due to rounding error as $i$ becomes close enough to $\pi/2$ that the ratio of $u$ and expression \eqref{eq:sec} equals $B q$, with $B$'s order of magnitude close to 1. This gives the following approximation of the angle at which the procedure stops working:
\begin{equation}
i_b = \arccos{\left[\left(\frac{B q}{4}\right)^{1/6}\right]},
\label{eq:sec2}
\end{equation}
where we set $\omega$ and $u'$ to 1 to obtain the worst-case scenario, since Equation \eqref{eq:sec} is then largest. When $i > i_b$, we set $i = \pi/2$, so that $u = u'$ and the sight line pierces the star if and only if $\tilde{y} \le \tilde{r}$. $B = 16$ and $q = 2\times10^{-16}$ yield $i_b = \pi / 2 - 0.0031$.

Once we have $\tilde{y}$, $u'$, $u$, and $\tilde{r}$ for a given sight line, we can compute $\tilde{z} = f\,u$ and
\begin{equation}
\cos{\phi} = \pm \sqrt{1 - \left(\frac{\tilde{y}}{\tilde{r}}\right)^2},
\end{equation}
where the negative sign corresponds to the case when the line goes through the upper half of the elliptical projection of its surface slice, i.e. when $u' > u \sin{i}$. Then, equations \eqref{eq:g} and \eqref{eq:T} produce gravity and temperature, which in turn allow us to compute $a_{ij}$ in Equation \eqref{eq:poly2} via the interpolation of Section \ref{subsec:interp}. At that point, equations \eqref{eq:poly}, \eqref{eq:mu}, and \eqref{eq:n} give us $I_\nu$ at the sight line.

\subsubsection{Blocked flux}

We wish to estimate $\mathcal{F}_{\nu,1}$, the flux that the planet blocks at a given time point. Multiplied by $D_{\star}^2$, it is equal to the integral of $I_\nu$ over the planet's circular projection, which is similar to the integral in equation \eqref{eq:int}, except that the two-dimensional integration domain is different. The first thought is to use a two-dimensional analogue of a numerical integration method similar to those in Section \ref{subsec:numerical}. For example, we could partition the integration domain into rectangles and employ a midpoint Riemann sum. However, since the integration domain is circular, we modify the latter method by attempting to partition the domain into circles instead. Specifically, let us say we can calculate $I_\nu$ at ${\cal N}_s$ sight lines within the domain. When ${\cal N}_s = 7$, we consider the problem of maximizing the total area of seven congruent, non-overlapping circles that fit within a circle of radius $R_1$; we call this the packing problem at ${\cal N}_s = 7$. Its solution, unique up to circular symmetry, is to place small circles of radius $R_1\,/\,3$ at the following locations in the $\mathbf{\hat{y}}\,\mathbf{\hat{z}'}$ coordinate system, with the system's origin shifted to match the center of the planet's projection \citep{graham1998dense}: 
\begin{equation}
\frac{R_1}{3}\,\{(0, 0), (\pm 2, 0), (\pm 1, \sqrt{3}), (\pm 1, -\sqrt{3})\}.
\label{eq:7pts}
\end{equation}
One of the small circles covers $1/9$ of the integration domain. We approximate its average intensity by $I_\nu$ at the circle's center. We then approximate the average intensity over the remaining $2/9$ of the domain by the average of all seven $I_\nu$ values. This latter average, multiplied by $\pi R_1^2$, is thus our estimate of $D_{\star}^2\mathcal{F}_{\nu,1}$. Given the known best solutions to the packing problem for $1 < {\cal N}_s < 19$, the proportion of the integration domain jointly covered by individual circles is highest for ${\cal N}_s = 7$, which thus gives the best estimate of the integral, by this measure. With ${\cal N}_s = 19$, we can use the packing in \citet{fodor_1999}, which covers more than $7/9$ of the domain. An alternative, with any ${\cal N}_s$, is picking points from a 2-dimensional distribution that is uniform over the projection. Each method averages $I_\nu$ over the representative sight lines and multiplies the result by the planet's projected area to obtain $D_{\star}^2\mathcal{F}_{\nu,1}$.

\begin{figure}[!t]
\includegraphics[width=\linewidth]{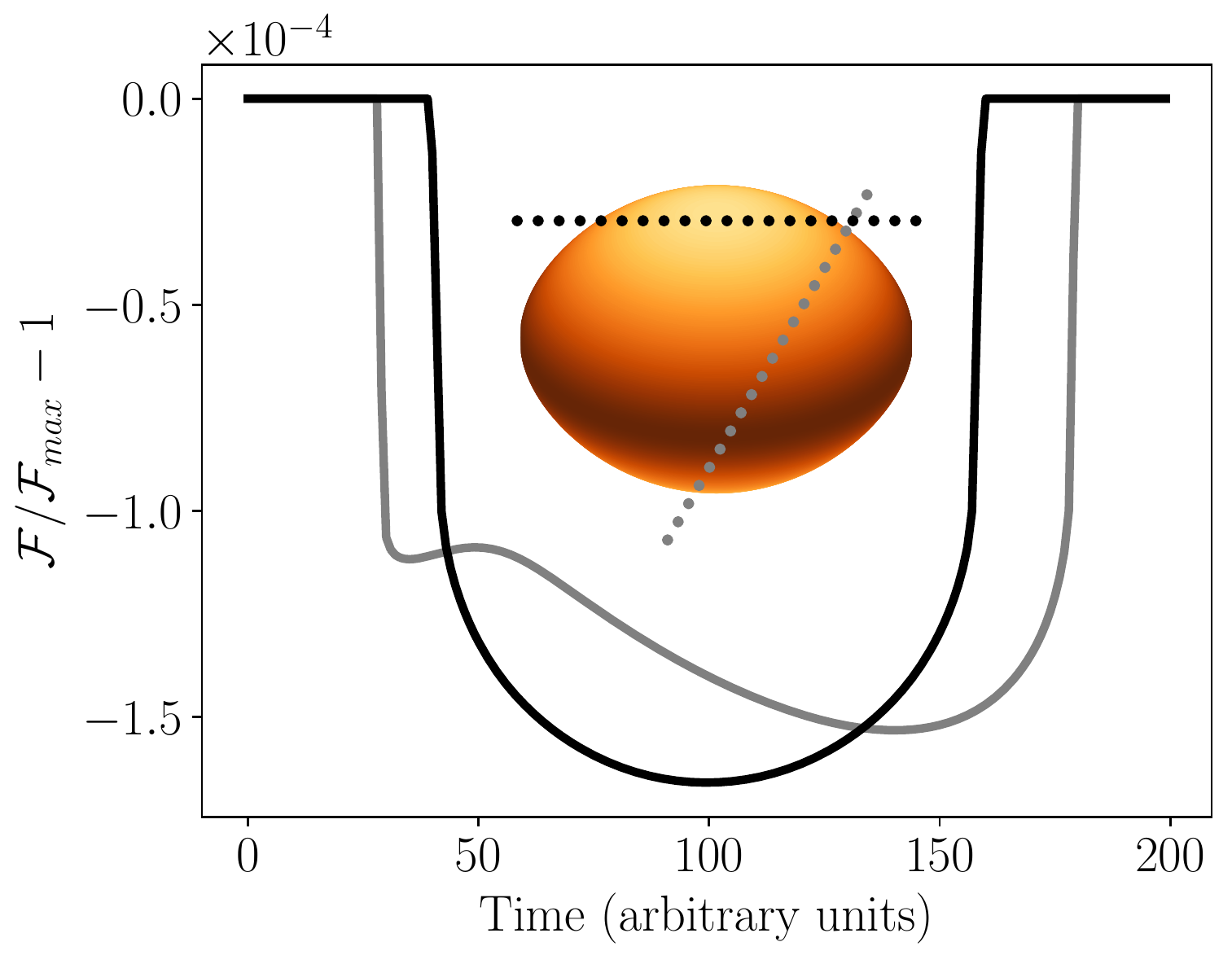}
\caption{Synthetic light curves in the Generic Bessel $V$ filter \citep{rodrigo_2012} of two separate transits by fictitious Jupiter-sized ($R_1 = 0.01 R_e$) planets orbiting Achernar. Inset: markers show the planets' progress from left to right, the orange color scheme traces local effective temperature on the stellar surface; each planet is enlarged for clarity. The black markers and line indicate a transit at $b = 0.6$ and $\alpha = 0$; grey symbols correspond to a transit at $b = -0.3$ and $\alpha = \pi / 3$ (see Section \ref{subsec:transit}). The blocked flux at each time point was computed from 7 packed-circle sight lines; the resulting light curves are indistinguishable from those based on about 100 random sight lines at each point.}
\label{fig:transit}
\end{figure}

We calculate the star's transit-free flux through the Generic Bessel $V$ filter via the right-hand side of equation \eqref{eq:filt} and call it ${\cal F}_{\rm max}$. At every time point, we use equations \eqref{eq:conv} and \eqref{eq:filt} to calculate blocked flux $\mathcal{F}_{1}$ through the same filter, with $\mathcal{F}_{\nu,1}$ instead of $\mathcal{F}_{\nu}$. Figure \ref{fig:transit} presents plots of 
\[
-\frac{\mathcal{F}_{1}}{{\cal F}_{\rm max}} = \frac{{\cal F} - {\cal F}_{\rm max}}{{\cal F}_{\rm max}}
\]
versus time, for two different systems. The plots resemble the curves predicted for similar systems in \citet{barnes_2009}. Here, ${\cal F}$ is the star's flux during transit. 

Computation of the transit curves in Figure \ref{fig:transit} starts with the star's inclination-independent calculation at ${\cal N} = 100$, which takes about 800 ms on a 2.3 GHz MacBook Pro with 8 GB of RAM (see the end of Section \ref{subsec:convergence}). Thereafter, calculation of blocked flux takes about 1.1 ms per sight line. The resulting total time for two transit curves with 200 time points each and 7 sight lines per time point is $\sim$4~seconds. 

In the software implementation of the integration scheme that we reference here \citep{lipatov2020}, we speed up the calculation of broadband photometry by moving the filtering step from the very end to the very beginning of such calculation. In this implementation, we replace ${\cal F_\nu}$ with $I_\nu$ and ${\cal F}$ with $I$ in equations \eqref{eq:conv} and \eqref{eq:filt}, use them to calculate $I$ for every $(T, g, \mu)$ grid point, obtain $I(\mu)$ fits for every $(T, g)$, perform the two-dimensional integration as before, and finally calibrate to the standard magnitude system via equation \eqref{eq:mag}. This change leads to a speed-up by a factor of about 5 when each star is treated at ten inclinations. For Generic Bessel B/V and HST ACS WFC F435W/F555W/F814W filters \citep{rodrigo_2012}, the associated errors in $I(\mu)$ fits do not exceed the equivalent $I_\nu(\mu)$ errors in Section \ref{subsec:fits}.

\section{Conclusion} \label{sec:concl}

We have presented  PARS  (Paint  the  Atmospheres  of  Rotating  Stars) -- a scheme for the integration of specific intensities over the surface of a rotating star to obtain the star's spectrum. Inputs to the scheme include the star's mass, luminosity, equatorial radius, rotational speed, and inclination.

We forgo differential rotation and volume-wide mass distribution in favor of solid-body rotation and a Roche model, respectively. This allows us to compute a closed-form expression for surface shape in cylindrical coordinates, based on \citet{EspinosaLara+Rieutord_2011}, a.k.a.~ER11. We then obtain closed-form expressions for the differential area element, cosine of the viewing angle $\mu$, and effective surface gravity $g$. We also adopt ER11's assumption that energy flux and gravity are collinear, which allows us to compute surface effective temperature $T$ with high precision up to 99.9\% of maximum rotation rate.

An important input to our scheme is a set of stellar atmosphere models \citep[such as in][a.k.a. CK04]{castelli_2004}, with intensity $I_\nu$ on a grid of $\mu$, $T$, and $g$. We model $I_\nu(\mu)$ as a piecewise 4th degree polynomial on a partition of $\mu$'s range and interpolate the polynomial coefficients in $\log{g}$ and the Planck function factor that involves $T$.

The polynomial form of $I_\nu(\mu)$, in combination with the closed-form expression for $\mu$, gives a closed-form expression for the indefinite integral in the azimuthal direction. The definite integral results from an algorithm that keeps track of the $\mu$ interval entered by the integration. We separately apply two cubic-fit numerical approximations to the longitudinal integrals below and above the equator, in view of abrupt intensity changes across this latitude at high rotational velocities.

Our scheme enables rapid calculation of synthetic spectra, taking just $\sim$2~seconds on a laptop computer to compute 1221-wavelength spectra at 50 inclinations. This, coupled with libraries of stellar models, will enable rigorous comparisons with data wherever the effects of stellar rotation are important. We highlight two examples: observed color-magnitude diagrams of star clusters and transit light curves. In future work we will further develop these applications. 

Individual spectral lines are not resolved with the atmosphere models we use in this article, though higher-resolution models can constitute input to PARS. On a related note, PARS does not generally account for the rotational Doppler effect. However, we do plan to utilize the scheme's framework to calculate the expected broadening of individual spectral lines in the future.

PARS Python source code is available for download and installation \citep{lipatov2020}. We have tested and run the software on a Unix operating system included with macOS 10. There is potential for porting PARS to other operating systems, as its core doesn't require the import of rare or highly specialized modules.

\acknowledgements{The authors would like to thank Nathan Bastian and Sebastian Kamann for comments on the manuscript. The authors also thank G.~Mirek Brandt for useful discussions in relation to this work.}

\software{PARS \citep{lipatov2020}, The NumPy Array \citep{numpy}, Matplotlib \citep{matplotlib}, SciPy \citep{scipy}.}

\clearpage

\appendix

\section{Piecewise integration}

\begin{algorithm}[H]
\caption{(see Section \ref{subsec:piecewise})}
\label{alg:piecewise}
\begin{algorithmic}[1]
\Require{
\[
p_{kj}(\tilde{z}, \phi) = 0 \;\; \mathrm{for} \;\; \{\tilde{z}, \phi\} \;\; \mathrm{s.t.} \;\; \mu(\tilde{z}, \phi) \notin m_j \equiv [\mu_j, \mu_{j+1}]
\]
}
\Ensure{
\[
P_{kj}(\tilde{z}) = \int^{\phi_b(\tilde{z})}_{0} \mu(\tilde{z}, \phi) \,\, p_{kj}\left(\tilde{z}, \phi\right)\,d\phi
\]
\begin{center}
is calculated $\forall \,\{k,j\}$ at a given $\tilde{z}$
\end{center}
}
\State $a \gets \sin{i}\,/\,n(\tilde{z})$ 
\State $b \gets -\cos{i}\,\,[f \times \tilde{r}'(\tilde{z})]\,/\,n(\tilde{z})$ 
\Function{$\mu$}{$\phi$} 
\Comment{$\mu$ as a function of $\phi$}
    \State \Return{ $a\,\cos{\phi} + b$ }
\EndFunction
\Function{$\phi$}{$\mu$} 
\Comment{$\phi$ as a function of $\mu$}
    \State \Return{ $\cos^{-1}{[\,(\,\mu - b\,)\,/\,a\,]}$ }
\EndFunction
\Procedure{int}{$\phi_a, \phi_b, m_j$} 
\Comment{integrate w.r.t. $\phi$ on a fixed $\mu$ interval}
    \State \textbf{require} $\mu(\phi) \in m_j \quad \forall \phi \in [\phi_a, \phi_b]$
    \State $P_{kj} \gets P_{kj} + \int^{\phi_b}_{\phi_a} \mu(\tilde{z}, \phi) \,\, p_{kj}\left(\tilde{z}, \phi\right)\,d\phi \quad \forall k$
\EndProcedure
\State \textbf{global} $P_{kj} \gets 0 \quad \forall\, \{k, j\}$ 
\Comment{the integrals we aim to compute}
\State $\phi_0 \gets 0$ 
\Comment{variable lower $\phi$ integration bound}
\State $\mu_u \gets \mu\left(\phi_0\right)$ 
\Comment{fixed upper $\mu$ integration bound}
\State $j \gets \max_{\mu_u \in m_{k}} k$ 
\Comment{index of the variable $\mu$ interval}
\If{$a \ne 0$} 
\Comment{if $\mu$ changes during integration}
    \If{$z < z_b$} 
    \Comment{if $\phi_b < \pi$}
        \State $\mu_l \gets 0$ 
        \Comment{the fixed lower $\mu$ integration bound is zero}
    \Else 
    \Comment{if $\phi_b = \pi$}
        \State $\mu_l \gets \mu(\pi)$ 
        \Comment{the fixed lower $\mu$ integration bound is above zero}
    \EndIf
    \While{$\mu_j > \mu_l$} 
    \Comment{while the $\mu$ interval lower bound is above the fixed lower $\mu$ integration bound}
        \State \Call{int}{$\phi_0,\, \phi(\mu_j),\,m_j$} 
        \Comment{integrate from the lower $\phi$ integration bound to $\phi$ at the lower bound of the $\mu$ interval}
        \State $\phi_0 \gets \phi(\mu_j)$ 
        \Comment{set the $\phi$ lower integration bound to $\phi$ corresponding to the lower bound of the $\mu$ interval}
        \State $j \gets j-1$ 
        \Comment{move to the next $\mu$ interval down}
    \EndWhile
    \State \Call{int}{$\phi_0,\,\phi(\mu_l),\,m_j$} 
    \Comment{integrate from the lower $\phi$ integration bound to $\phi$ at the fixed lower $\mu$ integration bound}
\Else 
\Comment{if $\mu$ is constant during integration}
    \State \Call{int}{$\phi_0,\,\pi,\,m_j$} 
    \Comment{integrate from the lower $\phi$ integration bound to $\pi$}
\EndIf
\end{algorithmic}
\end{algorithm}

\newpage

\bibliography{refs.bib}
\bibliographystyle{aasjournal}

\end{document}